\def\arxivversion{}
\documentclass{article}
\usepackage{iclr2027_conference,times}
\ifdefined\arxivversion
  \iclrfinalcopy
\fi
\usepackage[T1]{fontenc}
\usepackage[utf8]{inputenc}
\usepackage{amsmath,amssymb,amsthm,mathtools}
\usepackage{graphicx,booktabs,multirow}
\usepackage{pgfplots}
\pgfplotsset{compat=1.18}
\usepgfplotslibrary{groupplots}
\usetikzlibrary{positioning,shapes.geometric,fit}
\usepackage{hyperref}
\usepackage{cleveref}
\newtheorem{theorem}{Theorem}

\newtheorem{assumption}{Assumption}
\newcounter{algorithm}
\crefname{assumption}{Assumption}{Assumptions}
\Crefname{assumption}{Assumption}{Assumptions}
\newcommand{\E}{\mathbb E}
\newcommand{\Var}{\operatorname{Var}}
\newcommand{\Cov}{\operatorname{Cov}}

\newcommand{\ind}{\mathbf 1}
\newcommand{\DTE}{\Delta}
\title{Efficient Inference for\\Distributional Treatment Effects\\with Partially Observed Outcomes}
\ifdefined\arxivversion
\author{%
  Tomu Hirata\textsuperscript{1}\quad
  Undral Byambadalai\textsuperscript{2}\quad
  Tatsushi Oka\textsuperscript{3}\quad
  Shota Yasui\textsuperscript{2}\\
  \normalfont\textsuperscript{1}Individual Researcher\\
  \normalfont\textsuperscript{2}CyberAgent, Inc., Tokyo, Japan\\
  \normalfont\textsuperscript{3}Department of Economics, Keio University, Tokyo, Japan
}
\hypersetup{%
  hidelinks,
  pdftitle={Efficient Inference for Distributional Treatment Effects with Partially Observed Outcomes},
  pdfauthor={Tomu Hirata, Undral Byambadalai, Tatsushi Oka, Shota Yasui}
}

\else
  \author{Anonymous authors\\Paper under double-blind review}
\fi
\begin{document}
\maketitle
\ifdefined\arxivversion
  \lhead{Preprint}
\fi

\begin{abstract}
We study the distributional treatment effect when a surrogate outcome is fully observed, but the primary outcome is observed only for a subsample. This setting arises naturally in model or agent evaluation, where automated judges make it inexpensive to score every high-dimensional response while human ratings are often available for only a small subsample. When evaluation scores are discrete or ordinal, a difference between cumulative distribution functions (CDFs) directly measures changes in the probability of low-quality outputs at each rating cutpoint. Building on prediction-powered inference and surrogate-assisted treatment-effect estimation, we develop a cross-fitted estimator that combines conditional CDFs with and without surrogate information. We establish a Gaussian-process limit and uniform confidence bands over outcome thresholds, identify the semiparametric efficiency bound, and characterize variance gains over estimators that reweight the labeled subsample by its known sampling probabilities or adjust for covariates without the surrogate. The construction does not require continuous output density or agreement between the surrogate and primary-outcome scales. Through simulation results and four real data applications, we demonstrate the effectiveness of our approach.
\end{abstract}

\section{Introduction}
Randomized controlled trials (RCTs) provide a direct way to compare a new system with an existing one. In LLM applications, the intervention might be a change of model, prompt, or response-generation policy. Obtaining human ratings is costly: each output must be read and judged, sometimes by a subject matter expert. LLM judges can score responses at much larger scale \citep{zheng2023judging,kim2024prometheus}, making it practical to score every response automatically and obtain human ratings for a small subset. The challenge is to use automated scores to estimate an intervention's effect on \emph{human-rated quality}, even when the judge makes systematic errors.

A growing body of research uses predictions and imperfect annotations to support statistical inference. Prediction-powered inference (PPI), for example, combines predictions with a smaller labeled sample for inference on quantities such as means and quantiles \citep{angelopoulos2023prediction,angelopoulos2023ppipp}. The observed outcomes provide a correction for prediction errors. Related approaches include design-based inference with imperfect LLM annotations \citep{egami2023using} and surrogate-assisted treatment-effect estimation when primary outcomes are only partially observed \citep{kallus2025surrogates}. These methods provide a basis for using automated ratings without treating them as substitutes for human observations.

The choice of estimand also matters. Although the average treatment effect (ATE) is a common target, an average can conceal changes elsewhere in the outcome distribution \citep{jiang2023regression,byambadalai2024estimating}. Many LLM evaluations use \emph{ordinal ratings}: MT-Bench uses a 1--10 judge scale, Prometheus's Feedback Collection uses integer scores from 1 to 5, and HelpSteer2 provides human ratings from 0 to 4 \citep{zheng2023judging,kim2024prometheus,wang2024helpsteer2}. An evaluator may want to know whether an update reduces poor responses or increases the fraction meeting an acceptable-quality threshold. A \emph{distributional treatment effect} (DTE) answers this question by comparing the probability of a rating at or below a cutpoint between treatment arms. Taken over all nontrivial cutpoints, these contrasts describe the entire change in the ordinal distribution and are invariant to increasing recodings of the rating scale. Quantile treatment effects (QTEs) offer another view of distributional change, but regression-adjusted methods such as \citet{jiang2023regression} assume continuous outcomes. DTEs accommodate discrete and mixed outcomes directly. They can also reveal changes that a particular quantile misses: two arms may both have median rating~3 even if an update reduces the share rated at most~1 from 20\% to 10\%. The median effect is zero; the DTE at~1 records a 10-percentage-point reduction. No continuous outcome density is needed.

We develop a cross-fitted DTE estimator for experiments with sparsely observed primary outcomes, allowing labeling to depend on observed information and the surrogate and primary-outcome scales to differ. To quantify uncertainty across thresholds, we establish a functional central limit theorem and uniform confidence bands. The efficiency bound and variance decompositions characterize the precision available from surrogate information beyond covariate adjustment. We assess inferential validity in simulations with known truth, comparing against thresholdwise PPI++ and labeled-only regression adjustment. An offline randomized model comparison on BiGGen Bench then evaluates performance with human ratings, while HelpSteer2, wine reviews, and ACTG~175 provide further precision comparisons.

\section{Related Work}
\label{sec:related}
\paragraph{Prediction-powered and surrogate-assisted inference.}
Prediction-powered inference (PPI) combines a small labeled sample with predictions available for a larger sample, using the observed outcomes to correct prediction errors and construct confidence intervals for population quantities \citep{angelopoulos2023prediction}. PPI++ tunes the predictions' contribution for efficiency \citep{angelopoulos2023ppipp}; when predictors must be learned from the labeled sample, cross-PPI uses cross-fitting \citep{zrnic2024cross}. In LLM applications, design-based supervised learning corrects imperfect annotations \citep{egami2023using}, while hybrid evaluation combines prediction-powered inference with stratified designs \citep{boyeau2024autoeval,fisch2024stratified}. The statistical use of unlabeled observations has a longer history in semi-supervised inference \citep{chakrabortty2018efficient,zhang2019semi}. Recent work connects this literature to predictions as surrogate outcomes \citep{ji2025predictions} and general two-phase estimation \citep{kluger2026multiwave}.

The closest causal-inference foundation is \citet{kallus2025surrogates}, who derive efficient surrogate-assisted ATE estimation with limited outcome data. Applying a PPI mean procedure thresholdwise also yields distributional estimators, motivating our baselines. Our focus is joint distributional inference in one population under positive MAR labeling probabilities. This differs from transported DTE/QTE estimation \citep{wang2026transported} and long-term QTE data fusion \citep{liu2026double}, which connect source and target populations through transportability assumptions. We build on the pointwise foundations to establish a functional limit and uniform bands, with variance decompositions that isolate the surrogate's contribution.

\paragraph{Distributional and quantile treatment effects.}
Distributional estimations such as DTEs and QTEs describe treatment effects beyond the mean \citep{doksum1974empirical,lehmann1975nonparametrics,bitler2006mean}. Identification and testing have been studied for complier outcome distributions \citep{imbens1997distributions,abadie2002bootstrap} and counterfactual distributions under nonlinear difference-in-differences designs \citep{athey2006nonlinear}. For inference, \citet{firpo2007efficient} establishes efficient QTE estimation, while \citet{chernozhukov2013inference} develop functional-limit and bootstrap theory for counterfactual distributions. Distribution-function bands can also be used for quantile-effect inference with discrete outcomes \citep{chernozhukov2020generic}. High-dimensional and localized debiased methods accommodate flexible nuisance estimation \citep{belloni2017program,kallus2024localized}. Our labeled-only baseline follows the regression-adjusted DTE estimator of \citet{byambadalai2024estimating}. We extend the scope of distributional inference to partially observed primary outcomes, using fully observed surrogates to account for selective labeling and improve precision.

\section{Observation Model and Estimand}
\label{sec:setup}
For each unit, observe $O=(X,W,S,R,RY)$, where $X$ denotes pre-treatment covariates, $W\in\{0,1\}$ the evaluation arm, $S$ a fully observed auxiliary measurement (the surrogate), $Y$ the primary outcome of interest, and $R$ its observation indicator. The notation $RY$ is only an observation convention: no imputed value is treated as an observed primary outcome. We present two arms for clarity, and the construction applies to any pair of arms with overlap. We adopt the potential-outcomes framework \citep{rubin1974estimating}, with consistency $Y=Y(W)$. The surrogate may be affected by treatment, for example a judge score of the generated response, and may have a different scale from $Y$. This auxiliary role aligns with surrogate-assisted inference \citep{kallus2025surrogates}. We do not require classical surrogacy, $Y\perp W\mid(X,S)$, nor do we impose exchangeability on $S$, monotonicity, or instrumental-variable or shadow-variable restrictions. Treatment exchangeability and the propensity score condition on pre-treatment $X$ only, and the surrogate is not included in the treatment propensity score. The estimand concerns $Y$ alone. The relationship between $S$ and $Y$ is learned through a conditional CDF, under the MAR and positivity conditions in Assumption~\ref{ass:mar}.

Fix a nonrandom threshold set $\mathcal Y\subseteq\mathbb R$, which may be an interval, the whole line, or a finite grid $\{y_1<\cdots<y_J\}$. Our target is
\begin{equation}
 F_w(y)=P\{Y(w)\le y\},\qquad
 \DTE(y)=F_1(y)-F_0(y),\quad y\in\mathcal Y.
 \label{eq:target}
\end{equation}
For higher-is-better ratings, a negative contrast means fewer low-rated outcomes in arm~1. This is a contrast of marginal distributions, not the distribution of individual effects $Y(1)-Y(0)$.

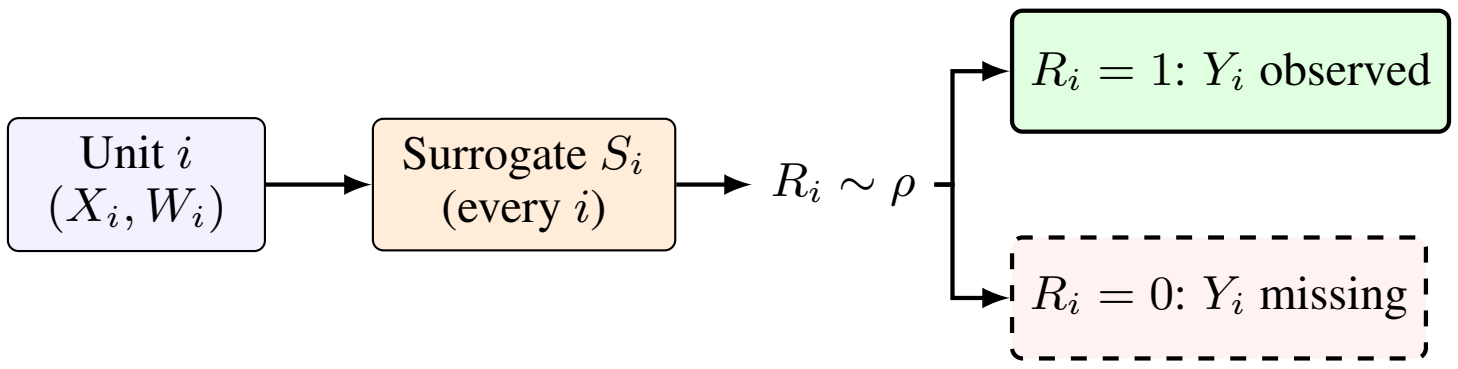
\begin{figure}[t]
\centering
\begin{tikzpicture}[
  font=\footnotesize,
  every node/.style={align=center},
  unit/.style={draw,rounded corners=2pt,minimum height=0.8cm,minimum width=1.7cm,fill=blue!5},
  sur/.style={draw,rounded corners=2pt,minimum height=0.8cm,minimum width=2cm,fill=orange!15},
  labeled/.style={draw,rounded corners=2pt,minimum height=0.8cm,minimum width=2.7cm,fill=green!12,thick},
  unlabeled/.style={draw,rounded corners=2pt,minimum height=0.8cm,minimum width=2.7cm,fill=red!5,dashed,thick},
  arr/.style={-latex,thick}
]
  \node[unit] (pop) {Unit $i$\\ $(X_i,W_i)$};
  \node[sur,right=0.7cm of pop] (sur) {Surrogate $S_i$\\ (every $i$)};
  \node[right=0.5cm of sur] (split) {$R_i\sim\rho$};
  \node[labeled,right=0.5cm of split,yshift=0.75cm] (lab) {$R_i=1$: $Y_i$ observed};
  \node[unlabeled,right=0.5cm of split,yshift=-0.75cm] (unl) {$R_i=0$: $Y_i$ missing};
  \draw[arr] (pop) -- (sur);
  \draw[arr] (sur) -- (split);
  \draw[arr] (split.east) -- ++(0.12,0) |- (lab.west);
  \draw[arr] (split.east) -- ++(0.12,0) |- (unl.west);
\end{tikzpicture}
\caption{Observation model. Every unit contributes covariates $X$, arm $W$, and a cheap surrogate $S$; the expensive primary outcome $Y$ is observed only on the labeled subsample ($R=1$), with labeling probability $\rho(X,W,S)$ (MAR, Assumption~\ref{ass:mar}). The proposed estimator combines both subsamples through a cross-fitted score that uses two conditional CDFs of the primary outcome, $\mu_w(y,X)$ and $\nu_w(y,X,S)$, together with the labeling and treatment probabilities.}
\label{fig:setup}
\end{figure}

\begin{assumption}[Sampling, exchangeability, and overlap]
\label{ass:rand}
The full unit records are i.i.d.; $Y(w)\perp W\mid X$ for each arm; and $\pi_w(X):=P(W=w\mid X)\ge\epsilon>0$ almost surely.
\end{assumption}
Independent randomized assignment implies exchangeability. In an observational application this assumption is substantive and cannot be verified by computing outcomes on a fully labeled sample.

\begin{assumption}[Label observability and selection]
\label{ass:mar}
The surrogate is observed for every unit, $R\perp Y\mid(X,W,S)$, and $\rho(X,W,S):=P(R=1\mid X,W,S)\ge\rho_{\min}>0$ almost surely.
\end{assumption}
Random annotation with a fixed positive sampling fraction is a special case. The more general MAR model permits oversampling particular surrogate scores, provided selection is independent of $Y$ conditional on $(X,W,S)$ and labeling probabilities remain positive. Accurate surrogate predictions are not required for identification. Their value for precision is quantified by \eqref{eq:surrogategain}, whose oracle variance gain is zero for a conditionally uninformative surrogate. Positivity is fixed as $n\to\infty$, so the results do not cover a labeling fraction vanishing with~$n$.

Define the two conditional CDFs of the primary outcome
\begin{equation}
 \mu_w(y,x)=P(Y\le y\mid X=x,W=w),\quad
 \nu_w(y,x,s)=P(Y\le y\mid X=x,W=w,S=s).
 \label{eq:nuisances}
\end{equation}
Under MAR, $\nu_w$ is identified from labeled observations conditional on $(X,W,S)$. Thus the surrogate can support identification under selective labeling as well as improve precision; it is not a treatment-adjustment covariate. The tower property and exchangeability then give
\begin{equation}
 F_w(y)=\E_X[\mu_w(y,X)]
 =\E_X\!\left[\E\{\nu_w(y,X,S)\mid X,W=w\}\right].
 \label{eq:id}
\end{equation}
The inner expectation averages over the surrogate distribution within each arm, rather than holding a post-treatment surrogate fixed. Thus, treatment-induced changes in the surrogate distribution are retained, and the target remains the total effect on the distribution of $Y$.
In empirical observational examples we use the standardized functional $\E_X[\mu_1(y,X)-\mu_0(y,X)]$ without claiming that exchangeability makes it causal. This functional is defined directly by the observed distribution and retains the same estimation formula under MAR.

\section{Surrogate-Augmented Distribution Regression}
\label{sec:method}
For $Z_y=\ind\{Y\le y\}$, define the score
\begin{align}
 \Psi_w(y;O,\eta)
 &=\mu_w(y,X)+\frac{\ind\{W=w\}}{\pi_w(X)}
       \{\nu_w(y,X,S)-\mu_w(y,X)\}\nonumber\\
 &\quad+\frac{\ind\{W=w\}R}{\pi_w(X)\rho(X,W,S)}
       \{Z_y-\nu_w(y,X,S)\}.
 \label{eq:score}
\end{align}
Here $\eta=(\mu,\nu,\pi,\rho)$ collects the nuisances. Only the last term uses observed primary outcomes. The first term standardizes over the target covariate distribution; the second integrates the surrogate-conditional prediction over the observed arm; the third corrects its prediction errors with labeled data. No comparison of $S$ and a threshold on the primary-outcome scale is needed.

Split the observations into a fixed number $K$ of approximately equal folds. Train all nuisances outside each evaluation fold and let $\widehat\eta^{-k(i)}$ denote the nuisance fits for unit~$i$. The estimator is
\begin{equation}
 \widehat F_w(y)=\frac1n\sum_{i=1}^n\widehat\Psi_{wi}(y),\qquad
 \widehat\DTE(y)=\widehat F_1(y)-\widehat F_0(y),\qquad
 \widehat\Psi_{wi}(y)=\Psi_w(y;O_i,\widehat\eta^{-k(i)}).
 \label{eq:estimate}
\end{equation}
When randomization or annotation probabilities are known by design, they can be inserted directly into the augmented score, eliminating the corresponding nuisance-estimation requirements. Cross-fitting avoids evaluating an observed primary outcome with a nuisance fit trained on that observation.

\begin{center}
\fbox{\begin{minipage}{\dimexpr\linewidth-2\fboxsep-2\fboxrule\relax}
\small
\refstepcounter{algorithm}\label{alg:dte}
\textbf{Algorithm \thealgorithm: Distributional inference with partially observed primary outcomes}

\textbf{Input:} $(X_i,W_i,S_i,R_i,R_iY_i)_{i=1}^n$, primary-outcome cutpoints, and level $1-\alpha$.
\begin{enumerate}\setlength{\itemsep}{1pt}\setlength{\parskip}{0pt}\setlength{\topsep}{2pt}
\item Split units into $K$ folds. Outside each evaluation fold, fit $\pi$ from $(X,W)$ and $\rho$ from $(X,W,S,R)$, or use supplied design probabilities.
\item For each arm and cutpoint, use labeled training units to regress $\ind\{Y\le y\}$ on $(X,S)$ for $\widehat\nu_w$, and on $X$ with weights $1/\widehat\rho$ for $\widehat\mu_w$.
\item Evaluate scores \eqref{eq:score} on held-out units. Pool all folds and average arm-score differences to obtain $\widehat\Delta(y)$ in \eqref{eq:estimate}.
\item Center these differences and compute SEs by \eqref{eq:se}. Use common multipliers across cutpoints to compute $T^{(b)}$, its empirical $(1-\alpha)$-quantile, and bands by \eqref{eq:band}.
\end{enumerate}
\textbf{Output:} DTE estimates and simultaneous bands over the evaluated cutpoints. Implementation safeguards are in Appendix~\ref{app:algorithm}.
\end{minipage}}
\end{center}

\paragraph{Estimating the outcome CDF conditional on covariates under MAR.}
An unweighted regression of $Z_y$ on $X$ in the labeled arm generally estimates $P(Y\le y\mid X,W=w,R=1)$, not $\mu_w$. We instead fit $\widehat\mu_w$ using inverse-label weights $1/\widehat\rho(X,w,S)$ on labeled training units. With the true weights,
\[
 \E\!\left[\frac R\rho\{Z_y-m(X)\}\,\middle|\,X,W=w\right]
 =\mu_w(y,X)-m(X).
\]
A sufficiently rich weighted regression can therefore recover $\mu_w$. Alternatively, regress a suitably cross-fitted estimate of $\nu_w$ on $X$ using \emph{all} training observations in arm~$w$. Algorithm~\ref{alg:dte} uses weighted regression for $\mu_w$ and unweighted labeled regression for $\nu_w$.

\paragraph{Robustness and nuisance learning.}
For fixed candidate nuisances $\bar\eta$ with positive weights, the score bias has the exact form
\begin{align}
 \E[\Psi_w(y;O,\bar\eta)]-F_w(y)
 &=\E\!\left[(\bar\mu_w-\mu_w)\frac{\bar\pi_w-\pi_w}{\bar\pi_w}\right]
 \nonumber\\[-2pt]
 &\quad+\E\!\left[\frac{\ind\{W=w\}}{\bar\pi_w}
 (\bar\nu_w-\nu_w)\frac{\bar\rho-\rho}{\bar\rho}\right].
 \label{eq:bias}
\end{align}
Arguments are suppressed for readability. This proves Neyman orthogonality and specifies the robustness combinations: each of the two products must vanish. For example, correct design weights make the score unbiased for any fixed outcome regressions, and correct outcome regressions make it unbiased for any positive candidate weights. Efficiency additionally requires learning the true conditional CDFs. Arbitrary surrogate quality is permitted.

\section{Uniform Inference and Efficiency}
\label{sec:theory}
Write $\varphi_w(y;O)=\Psi_w(y;O,\eta)-F_w(y)$ and $\varphi=\varphi_1-\varphi_0$. Norms integrate over a fresh $O\sim P$, holding the training sample fixed; fitted-function norms are therefore random. Let $\mathbb G_n f=\sqrt n(P_n-P)f$ and $\Omega(y,y')=\E[\varphi(y;O)\varphi(y';O)]$.

\begin{assumption}[Uniform nuisance rates and score complexity]
\label{ass:rates}
For every fold and arm, fitted CDF values lie in $[0,1]$, $\widehat\pi_w\ge\epsilon/2$, and $\widehat\rho\ge\rho_{\min}/2$, with fitted probabilities at most one. All nuisances are $L^2(P)$ consistent uniformly over $\mathcal Y$, and
\begin{equation}
 \sup_{y\in\mathcal Y}\|\widehat\mu_w-\mu_w\|_2\|\widehat\pi_w-\pi_w\|_2
 +\sup_{y\in\mathcal Y}\|\widehat\nu_w-\nu_w\|_2\|\widehat\rho-\rho\|_2
 =o_P(n^{-1/2}).
 \label{eq:rates}
\end{equation}
The number of folds is fixed, with positive limiting fold proportions. Conditional on each training sample, the fitted score class $\widehat{\mathcal F}_{kw}=\{\Psi_w(y;\cdot,\widehat\eta^{-k}):y\in\mathcal Y\}$ is pointwise measurable and, with probability tending to one, satisfies
\begin{equation}
 \sup_Q N(uM,\widehat{\mathcal F}_{kw},L^2(Q))\le(A/u)^v,
 \qquad 0<u\le1,
 \label{eq:entropy}
\end{equation}
for fixed constants $A\ge e,v,M<\infty$, with $M$ also a constant envelope; $Q$ ranges over finitely supported probability measures. The indexed processes are separable, with measurable suprema.
\end{assumption}
Rates $o_P(n^{-1/4})$ suffice, but known weights remove the corresponding product restrictions. The entropy condition is automatic on a fixed finite grid and holds for fixed-dimensional logistic score classes; it is not a consistency guarantee for misspecified learners. Appendix~\ref{app:joint} verifies these complexity claims and proves the true-score class is Donsker.

\begin{theorem}[Gaussian-process limit and uniform bands]
\label{thm:joint}
Under Assumptions~\ref{ass:rand}--\ref{ass:rates},
\[
 \sqrt n(\widehat\Delta-\Delta)=\mathbb G_n\varphi+o_P(1)
 \rightsquigarrow\mathbb G\quad\text{in }\ell^\infty(\mathcal Y),
\]
where the remainder vanishes in supremum norm and $\mathbb G$ is a tight centered Gaussian process with covariance kernel $\Omega$. The fitted covariance kernel is uniformly consistent, and the Rademacher multiplier process below converges conditionally to $\mathbb G$. On a fixed nonempty set $\mathcal Y_0\subseteq\mathcal Y$ with $\inf_{y\in\mathcal Y_0}\Omega(y,y)>0$, its studentized bands have asymptotic simultaneous coverage $1-\alpha$, provided the limiting supremum distribution is continuous and strictly increasing at its $(1-\alpha)$-quantile.
\end{theorem}
Let $\widehat\varphi_i(y)=\widehat\Psi_{1i}(y)-\widehat\Psi_{0i}(y)-\widehat\Delta(y)$. Set
\begin{equation}
 \widehat\Omega(y,y')=\frac1n\sum_i\widehat\varphi_i(y)\widehat\varphi_i(y'),
 \qquad \widehat{\mathrm{SE}}(y)=\sqrt{\widehat\Omega(y,y)/n}.
 \label{eq:se}
\end{equation}
Taking the arm-score difference retains their covariance. For independent Rademacher multipliers $\zeta_i^{(b)}$, independent of the data, compute
\begin{equation}
 T^{(b)}=\sup_{y\in\mathcal Y_0}\left|
 \frac{n^{-1/2}\sum_i\zeta_i^{(b)}\widehat\varphi_i(y)}
      {\sqrt{\widehat\Omega(y,y)}}\right|,
 \qquad
 C_n(y)=\left[\widehat\Delta(y)\pm\widehat c_{1-\alpha}\widehat{\mathrm{SE}}(y)\right],
 \label{eq:band}
\end{equation}
Here $\widehat c_{1-\alpha}$ is the conditional $(1-\alpha)$-quantile of $T^{(b)}$:
\[
 \widehat c_{1-\alpha}=\inf\{t:P_*(T^{(b)}\le t)\ge1-\alpha\},
\]
where $P_*$ holds the observed data fixed and randomizes only the multipliers. In computation, we approximate this quantile by the empirical $(1-\alpha)$-quantile of $T^{(1)},\ldots,T^{(B)}$ from $B$ independent multiplier draws. The same critical value is used at every threshold in $C_n(y)$; pointwise intervals instead use $z_{1-\alpha/2}$. The unstudentized process limit permits $\mathcal Y=\mathbb R$; studentization excludes thresholds with vanishing variance. Unstudentized bands are given in Appendix~\ref{app:joint}.

\paragraph{Ordinal outcomes and finite grids.}
For an outcome with a fixed finite number of ordered categories, all nontrivial cutpoints form a finite set and determine the full CDF contrast. The entropy condition in Assumption~\ref{ass:rates} is then automatic; the nuisance consistency and product-rate requirements remain. On $\mathcal Y_0=\{y_1,\ldots,y_J\}$, the limit is $N_J(0,(\Omega(y_j,y_\ell))_{j\ell})$ and the supremum is a maximum. Positive coordinate variances suffice for quantile continuity, even with singular covariance, so Theorem~\ref{thm:joint} gives simultaneous coverage of the full ordinal CDF contrast under its remaining assumptions. Algorithm~\ref{alg:dte} fits binary outcome regressions at each cutpoint, using $X$ for $\mu_w$ and $(X,S)$ for $\nu_w$; a 1--5 judge score can inform a 0--4 human-rating scale without a separate calibration map. For continuous outcomes, evaluating a fixed grid establishes grid coverage; approximating a continuum supremum requires separate discretization control.

\begin{theorem}[Semiparametric efficiency bound and oracle IPW comparison]
\label{thm:efficiency}
For any fixed finite collection $y_1,\ldots,y_J\in\mathcal Y$, in the nonparametric observed-data model satisfying Assumptions~\ref{ass:rand}--\ref{ass:mar}, $\boldsymbol\varphi(O)=(\varphi(y_j;O))_{j=1}^J$ is the efficient influence function of the DTE vector. Its semiparametric efficiency bound, with $\Omega_{j\ell}=\Omega(y_j,y_\ell)$, is
\[
 \Omega=\E[\boldsymbol\varphi(O)\boldsymbol\varphi(O)^\top],
\]
which lower-bounds the limiting covariance of regular estimators scaled by $\sqrt n$. Under Assumption~\ref{ass:rates}, our estimator attains this bound, giving efficient asymptotic variance $\Omega_{jj}/n$ at cutpoint $y_j$.
For oracle IPW, with arm score $H_w=\ind\{W=w\}RZ_y/(\pi_w\rho)$,
\begin{equation}
 \Var(H_1-H_0)=\Var\{\varphi(y;O)\}
 +\Var(A_1-A_0+B_1-B_0),
 \label{eq:ipwgain}
\end{equation}
where $A_w=\mu_w(\ind\{W=w\}/\pi_w-1)$ and
$B_w=\nu_w\ind\{W=w\}(R/\rho-1)/\pi_w$.
\end{theorem}
Thus the feasible efficient estimator asymptotically weakly improves on \emph{known-weight} unadjusted IPW. Plug-in IPW with estimated weights has additional first-order terms and need not have the oracle variance \citep{hirano2003efficient}. Its uncertainty must be computed for its actual fitting procedure. The comparison in \eqref{eq:ipwgain} does not transfer by asserting equality of feasible and oracle IPW variances.

\paragraph{Isolating the surrogate's contribution.}
Part of the gain against unnormalized IPW is available even without a useful surrogate. A more informative comparator uses the same labeling weights and a covariate-only outcome regression:
\[
 C_w=\mu_w(y,X)+\frac{\ind\{W=w\}R}{\pi_w\rho}\{Z_y-\mu_w(y,X)\}.
\]
An orthogonal decomposition gives
\begin{equation}
 \Var(C_1-C_0)-\Var\{\varphi(y;O)\}
 =\sum_{w=0}^1\E\!\left[
 \frac{\ind\{W=w\}}{\pi_w^2}\frac{1-\rho}{\rho}
 \{\nu_w(y,X,S)-\mu_w(y,X)\}^2\right].
 \label{eq:surrogategain}
\end{equation}
This is strictly positive when the surrogate changes a conditional cutpoint probability on a set with missing labels of positive probability. It vanishes for a conditionally uninformative surrogate or fully observed primary outcomes. Under MAR the comparator may still use the surrogate in its labeling model; ``covariate-only'' describes its outcome regression.

\section{Experiments}
\label{sec:experiments}
We first assess inferential validity in simulations with known population truth. We then compare empirical precision in BiGGen Bench, HelpSteer2, wine, and ACTG~175 under their stated sampling and modeling assumptions. Additional BiGGen repetitions assess error against its fully rated prompt pool. Benchmark coverage and application SEs have different interpretations from population coverage in the simulations; implementation details are in Appendix~\ref{app:experiments}.

\subsection{Simulation: continuous and ordinal primary outcomes}
\label{sec:simulation}
We generate $X\sim N(0,I_4)$, independent $W\sim\operatorname{Bernoulli}(1/2)$, and
\[
 Y(0)=0.3X_1-0.2X_2+\varepsilon,\qquad
 Y(1)=0.6X_1+0.4X_2^2+\varepsilon,\qquad \varepsilon\sim N(0,0.5^2).
\]
The continuous surrogate is $S=Y+0.3(X_1-0.5X_2)+\eta$, with independent $\eta\sim N(0,0.6^2)$. The arms differ in shape as well as location. In the continuous experiment, labeling depends on covariates and the surrogate through a fixed bounded MAR mechanism, with expected labeled fraction $0.2$. In the ordinal experiment, both the primary outcome and surrogate are discretized into five categories using boundaries $(-0.75,-0.25,0.25,0.75)$ and labeling is MCAR with probability $0.2$. The latter mirrors a five-level judge rubric. We evaluate continuous thresholds $\{-0.75,-0.5,0,0.5,1,1.5\}$ and ordinal cutpoints $\{0,1,2,3\}$. The continuous grid starts near the 10th percentile in each arm. The true DTE is evaluated by one-dimensional numerical integration, avoiding Monte Carlo error in the reference CDF.

We use five-fold cross-fitting and logistic distribution regression, including $X_2^2$ among covariates. Comparators are known-weight IPW, the covariate-only outcome-adjustment ablation, labeled-only empirical CDFs, and surrogate-only CDFs. The ordinal MCAR experiment adds untuned thresholdwise PPI, PPI++ \citep{angelopoulos2023ppipp} with raw-score and learned-CDF predictions, and labeled-only regression-adjusted DTE estimation \citep{byambadalai2024estimating}. Labeled-only RA uses labeled records for both fitting and averaging; PPI++ uses disjoint labeled and unlabeled arm samples with training-only tuning. These methods are compared under randomized MCAR sampling, where they identify the same target; they are not applied unchanged to MAR or observational data. Appendix~\ref{app:baseline-methods} specifies the implementations.

\begin{figure}[t]
 \centering
\definecolor{cproposed}{HTML}{1670b5}
\definecolor{cnosurrogate}{HTML}{c88400}
\definecolor{cppipp}{HTML}{cc6677}
\definecolor{cppipplearned}{HTML}{332288}
\definecolor{cgoldra}{HTML}{117733}
\definecolor{cppi}{HTML}{555555}
\definecolor{cgold}{HTML}{259642}
\definecolor{cipw}{HTML}{8b4fa6}
\definecolor{csurrogate}{HTML}{d24b40}
\begin{tikzpicture}
\begin{groupplot}[group style={group size=3 by 2,horizontal sep=.95cm,vertical sep=.65cm},
scale only axis,width=.25\linewidth,height=2.85cm,
tick label style={font=\fontsize{8}{9}\selectfont},
label style={font=\fontsize{9}{10}\selectfont},title style={font=\small},
xlabel={Outcome cutpoint},grid=major,grid style={gray!15},
scaled ticks=false,yticklabel style={/pgf/number format/fixed,/pgf/number format/precision=3},axis line style={gray!60},
tick align=outside,major tick length=2pt,every axis plot/.append style={mark size=1.5pt},
legend style={font=\fontsize{8}{10}\selectfont,draw=none,column sep=5pt,row sep=2pt},legend columns=3]
\nextgroupplot[xlabel={},title={Bias ($\times10^{-2}$)},ylabel={Continuous (MAR)}]
\addplot+[color=cproposed,mark=*,solid,line width=1pt] coordinates {(-0.75,-0.152206) (-0.5,-0.2829245) (0,0.1532372) (0.5,0.1418378) (1,-0.1434907) (1.5,-0.2212144)};
\addplot+[color=cnosurrogate,mark=square*,dashed,line width=.7pt] coordinates {(-0.75,-0.1466406) (-0.5,-0.2742808) (0,0.1463567) (0.5,0.145659) (1,-0.1490674) (1.5,-0.2078312)};
\addplot+[color=cipw,mark=triangle,solid,line width=.7pt] coordinates {(-0.75,0.008894524) (-0.5,-0.157704) (0,0.1479195) (0.5,0.07644416) (1,-0.09205343) (1.5,-0.1065819)};
\addplot[black,densely dotted,forget plot] coordinates {(-0.75,0) (1.5,0)};
\nextgroupplot[xlabel={},title={95\% coverage},ymin=.80,ymax=1,ytick={.80,.85,.90,.95,1}]
\addplot+[color=cproposed,mark=*,solid,line width=1pt] coordinates {(-0.75,0.95) (-0.5,0.944) (0,0.952) (0.5,0.944) (1,0.954) (1.5,0.942)};
\addplot+[color=cnosurrogate,mark=square*,dashed,line width=.7pt] coordinates {(-0.75,0.944) (-0.5,0.95) (0,0.94) (0.5,0.958) (1,0.958) (1.5,0.948)};
\addplot+[color=cipw,mark=triangle,solid,line width=.7pt] coordinates {(-0.75,0.944) (-0.5,0.938) (0,0.938) (0.5,0.944) (1,0.948) (1.5,0.936)};
\addplot[black,densely dotted,forget plot] coordinates {(-0.75,0.95) (1.5,0.95)};
\nextgroupplot[xlabel={},title={MSE ($\times10^{-3}$)},ymode=log,log basis y=10,ytick={0.1,0.2,0.5,1,2,5,10},yticklabels={0.1,0.2,0.5,1,2,5,10},minor y tick num=0]
\addplot+[color=cproposed,mark=*,solid,line width=1pt] coordinates {(-0.75,0.8038255) (-0.5,1.12836) (0,1.437187) (0.5,1.083783) (1,0.5405167) (1.5,0.2509869)};
\addplot+[color=cnosurrogate,mark=square*,dashed,line width=.7pt] coordinates {(-0.75,1.004812) (-0.5,1.398003) (0,1.843509) (0.5,1.316555) (1,0.6477902) (1.5,0.2914758)};
\addplot+[color=cipw,mark=triangle,solid,line width=.7pt] coordinates {(-0.75,1.486366) (-0.5,2.5065) (0,5.480152) (0.5,7.838987) (1,9.845616) (1.5,10.78211)};
\nextgroupplot[ylabel={Ordinal (MCAR)},xtick={0,1,2,3}]
\addplot+[color=cproposed,mark=*,solid,line width=1pt] coordinates {(0,-0.1291362) (1,-0.06304077) (2,0.2138426) (3,0.0347785)};
\addplot+[color=cnosurrogate,mark=square*,dashed,line width=.7pt] coordinates {(0,-0.1207402) (1,-0.04633294) (2,0.1819485) (3,0.007661234)};
\addplot+[color=cppipp,mark=triangle*,dashed,line width=.8pt] coordinates {(0,-0.1232999) (1,-0.01722788) (2,0.06020912) (3,-0.1172442)};
\addplot+[color=cppipplearned,mark=diamond*,densely dotted,line width=.9pt] coordinates {(0,-0.1311657) (1,-0.0409518) (2,0.2069565) (3,-0.04855003)};
\addplot+[color=cgoldra,mark=square,dashdotted,line width=.8pt] coordinates {(0,-0.1041325) (1,-0.08134583) (2,0.06217957) (3,-0.09603943)};
\addplot+[color=cipw,mark=triangle,solid,line width=.7pt] coordinates {(0,-0.1222806) (1,-0.1355205) (2,-0.02811452) (3,-0.119751)};
\addplot[black,densely dotted,forget plot] coordinates {(0,0) (3,0)};
\nextgroupplot[ymin=.80,ymax=1,ytick={.80,.85,.90,.95,1},xtick={0,1,2,3}]
\addplot+[color=cproposed,mark=*,solid,line width=1pt] coordinates {(0,0.958) (1,0.962) (2,0.962) (3,0.958)};
\addplot+[color=cnosurrogate,mark=square*,dashed,line width=.7pt] coordinates {(0,0.954) (1,0.944) (2,0.95) (3,0.96)};
\addplot+[color=cppipp,mark=triangle*,dashed,line width=.8pt] coordinates {(0,0.944) (1,0.958) (2,0.938) (3,0.946)};
\addplot+[color=cppipplearned,mark=diamond*,densely dotted,line width=.9pt] coordinates {(0,0.952) (1,0.96) (2,0.946) (3,0.954)};
\addplot+[color=cgoldra,mark=square,dashdotted,line width=.8pt] coordinates {(0,0.96) (1,0.936) (2,0.958) (3,0.946)};
\addplot+[color=cipw,mark=triangle,solid,line width=.7pt] coordinates {(0,0.946) (1,0.938) (2,0.938) (3,0.95)};
\addplot[black,densely dotted,forget plot] coordinates {(0,0.95) (3,0.95)};
\nextgroupplot[ymode=log,log basis y=10,ytick={0.1,0.2,0.5,1,2,5,10},yticklabels={0.1,0.2,0.5,1,2,5,10},minor y tick num=0,xtick={0,1,2,3},legend to name=simulationlegend]
\addplot+[color=cproposed,mark=*,solid,line width=1pt] coordinates {(0,0.6630871) (1,1.305217) (2,1.354741) (3,0.9482396)};
\addlegendentry{Proposed}
\addplot+[color=cnosurrogate,mark=square*,dashed,line width=.7pt] coordinates {(0,0.7548049) (1,1.655757) (2,1.597458) (3,1.026582)};
\addlegendentry{Covariate-only}
\addplot+[color=cppipp,mark=triangle*,dashed,line width=.8pt] coordinates {(0,0.7317318) (1,1.511614) (2,1.767447) (3,1.137986)};
\addlegendentry{PPI++ (raw)}
\addplot+[color=cppipplearned,mark=diamond*,densely dotted,line width=.9pt] coordinates {(0,0.6881827) (1,1.385196) (2,1.572973) (3,1.111198)};
\addlegendentry{PPI++ (CDF)}
\addplot+[color=cgoldra,mark=square,dashdotted,line width=.8pt] coordinates {(0,0.8141507) (1,1.727652) (2,1.642697) (3,1.176387)};
\addlegendentry{Labeled-only RA}
\addplot+[color=cipw,mark=triangle,solid,line width=.7pt] coordinates {(0,1.054664) (1,3.17166) (2,5.794219) (3,7.879752)};
\addlegendentry{Oracle IPW}
\end{groupplot}
\node[anchor=north] at ([yshift=-1.05cm]group c2r2.south) {\pgfplotslegendfromname{simulationlegend}};
\end{tikzpicture}
 \caption{Known-truth simulations at $n=2000$ with 500 repetitions. Top: a continuous primary outcome with MAR labeling. Bottom: a five-category ordinal primary outcome with MCAR labeling. The panels report bias, pointwise 95\% coverage, and MSE. MSE uses a logarithmic vertical axis, with tick values in units of $10^{-3}$. The ordinal panels include raw-score PPI++, learned-CDF PPI++, and labeled-only RA; Table~\ref{tab:ordinal-baselines} also reports the simpler baselines.}
 \label{fig:simulation}
\end{figure}
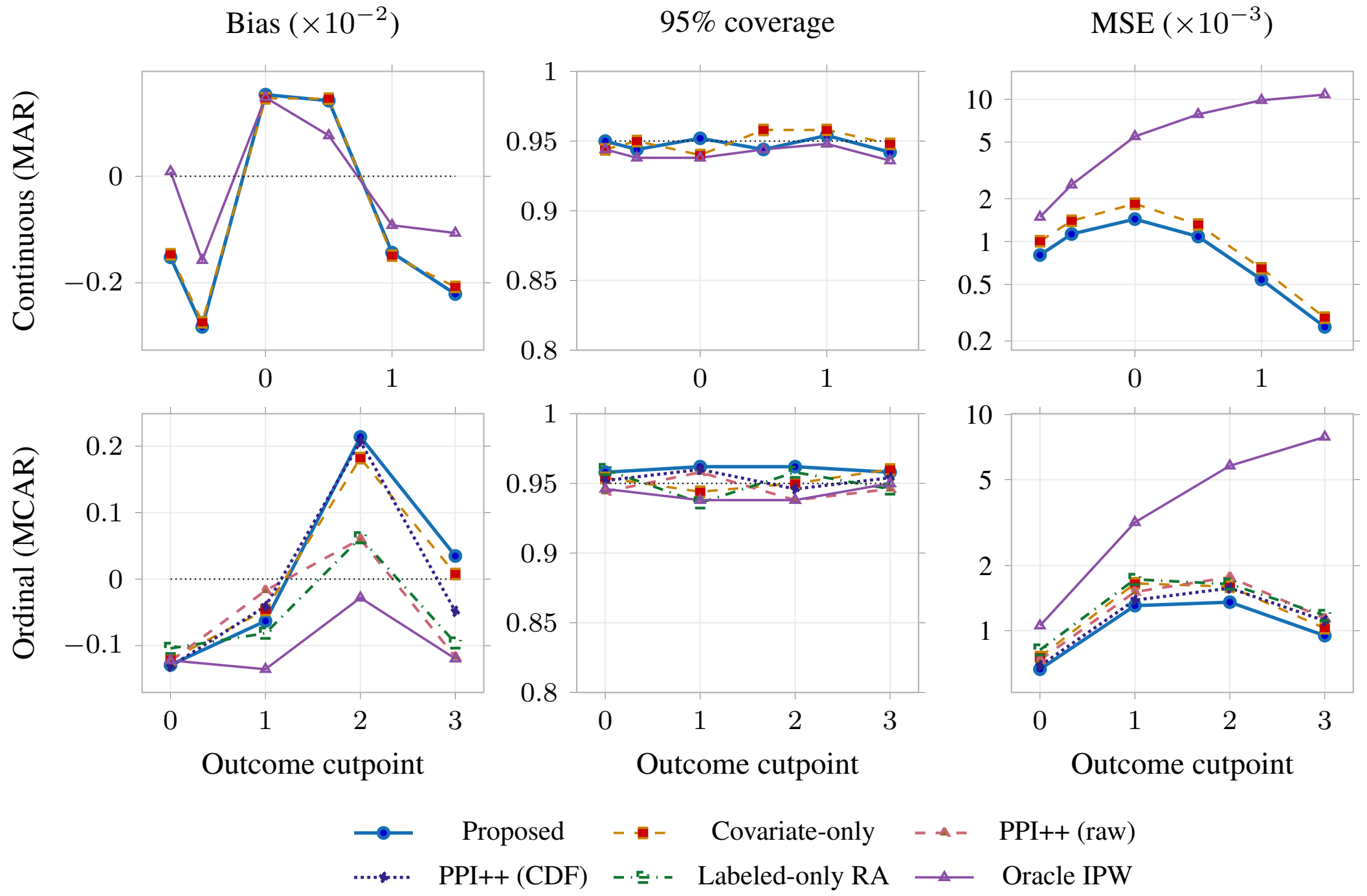

\paragraph{Results.} Figure~\ref{fig:simulation} shows that, in the ordinal experiment, the proposed method reduces MSE across cutpoints by 30--45\% relative to labeled-only, 34--48\% relative to untuned PPI, and 8--21\% relative to covariate-only adjustment. Its pointwise coverage ranges from $0.958$ to $0.962$. Against the stronger baselines in Table~\ref{tab:ordinal-baselines}, the reductions are 9--23\% relative to raw-score PPI++, 4--15\% relative to learned-CDF PPI++, and 18--24\% relative to labeled-only RA. For continuous primary outcomes under MAR, MSE reductions across cutpoints are 46--98\% relative to oracle IPW and 14--22\% relative to covariate-only adjustment. Continuous pointwise coverage ranges from $0.942$ to $0.954$.

For each repetition, we approximate $\widehat c_{0.95}$ with 500 Rademacher realizations of $T^{(b)}$ in \eqref{eq:band}. With $n=2000$, 5000 repetitions per outcome, and both $\pi$ and $\rho$ estimated, nominal 95\% bands attain simultaneous coverage over all evaluated cutpoints of $0.951$ for ordinal outcomes and $0.938$ for continuous outcomes. The mean ratios of band to pointwise-interval widths are $1.26$ and $1.32$, respectively.

\subsection{Applications}
\label{sec:empirical}
\paragraph{BiGGen Bench.}
\label{sec:biggen}
We compare Mistral-7B-Instruct-v0.2 with Mixtral-8x7B-Instruct-v0.1 on 695 shared prompts from BiGGen Bench \citep{kim2025biggen}. Each prompt is assigned one model's released response with probability $1/2$; exactly 200 responses are sampled uniformly for human rating. The primary outcome is the human 1--5 task-specific rubric rating, and the surrogate is the cached GPT-4 Turbo score. We adjust for user- and system-prompt lengths, character diversity, and capability indicators. The DTE measures how switching to Mixtral changes the probability of a rating at or below each cutpoint, without assuming equal distances between rating categories. The application uses the first archived assignment and annotation sample, with every nontrivial cutpoint $y\in\{1,2,3,4\}$.

\paragraph{HelpSteer2.}
We use 20,324 prompt--response pairs with human helpfulness ratings from HelpSteer2 \citep{wang2024helpsteer2}. The contrast compares released complexity ratings at least~2 against simpler responses, adjusting for prompt length and character diversity. Human ratings and cached judge scores use a 0--4 scale, with 399 human ratings retained under a 2\% MCAR mask. The judge applies the helpfulness rubric without being fitted to the released human ratings. Cutpoint contrasts describe how the frequency of low helpfulness ratings differs by response complexity.

\paragraph{Wine reviews.}
We compare red and white varieties among 10,000 US wine reviews, adjusting for log price, its square, and state \citep{ji2025predictions}. The primary outcome is the professional points rating; the surrogate is an LLM score of the tasting description on a 1--5 scale. This application illustrates unequal primary-outcome and surrogate scales: the outcome regressions predict primary-outcome cutpoint probabilities directly from the auxiliary rating. A 20\% MCAR mask retains 2,049 professional ratings.

\paragraph{ACTG 175.}
We compare zidovudine--didanosine with zidovudine monotherapy \citep{hammer1996trial} in 1,054 patients, using week-20 CD4 as a surrogate for week-96 CD4, observed for 654 patients. Here missing primary outcomes arise in follow-up rather than from an imposed annotation mask. We adjust for fourteen baseline covariates, including baseline CD4, age, and weight, and fit a labeling score using these covariates, treatment, and week-20 CD4.

\paragraph{Uncertainty and results.}
Table~\ref{tab:empirical} reports each cutpoint separately. Every SE is the standard deviation of 200 record-bootstrap estimates, refitting nuisances and repeating cross-fitting with common resamples across methods. Assignment and observed labels are fixed before bootstrapping. These are precision diagnostics, without corrections for additional prompt, reviewer, finite-population, or trial-design dependence.
\begin{table}[t]
\centering\small
\begin{tabular}{llrrrrrrr}
\toprule
& & & \multicolumn{3}{c}{Bootstrap standard error} & \multicolumn{2}{c}{SE ratio} & MSE ratio \\
\cmidrule(lr){4-6}\cmidrule(lr){7-8}\cmidrule(lr){9-9}
Dataset & $y$ & $\widehat\Delta$ & Proposed & Covariate & IPW & / IPW & / Cov. & / Cov. \\
\midrule
BiGGen & 1 & $+0.025$ & 0.042 & 0.045 & 0.045 & 0.95 & 0.95 & 0.80 \\
 & 2 & $+0.012$ & 0.055 & 0.056 & 0.063 & 0.87 & 0.99 & 0.80 \\
 & 3 & $-0.058$ & 0.066 & 0.064 & 0.080 & 0.82 & 1.03 & 0.78 \\
 & 4 & $-0.052$ & 0.068 & 0.069 & 0.112 & 0.61 & 0.99 & 0.90 \\
\midrule
HelpSteer2 & 0 & $-0.054$ & 0.027 & 0.030 & 0.032 & 0.85 & 0.91 & --- \\
 & 1 & $-0.096$ & 0.036 & 0.041 & 0.046 & 0.80 & 0.89 & --- \\
 & 2 & $-0.193$ & 0.047 & 0.049 & 0.055 & 0.85 & 0.96 & --- \\
 & 3 & $-0.037$ & 0.050 & 0.052 & 0.073 & 0.69 & 0.97 & --- \\
\midrule
Wine & 85 & $+0.073$ & 0.012 & 0.017 & 0.021 & 0.58 & 0.72 & --- \\
 & 87 & $+0.106$ & 0.018 & 0.022 & 0.030 & 0.59 & 0.81 & --- \\
 & 89 & $+0.098$ & 0.019 & 0.022 & 0.037 & 0.52 & 0.87 & --- \\
 & 91 & $+0.068$ & 0.020 & 0.021 & 0.044 & 0.46 & 0.96 & --- \\
 & 93 & $+0.062$ & 0.018 & 0.018 & 0.050 & 0.36 & 1.00 & --- \\
\midrule
ACTG 175 & 100 & $-0.107$ & 0.027 & 0.027 & 0.032 & 0.86 & 1.00 & --- \\
 & 200 & $-0.173$ & 0.032 & 0.033 & 0.042 & 0.75 & 0.97 & --- \\
 & 300 & $-0.153$ & 0.033 & 0.035 & 0.052 & 0.64 & 0.95 & --- \\
 & 400 & $-0.119$ & 0.032 & 0.032 & 0.058 & 0.55 & 0.99 & --- \\
 & 500 & $-0.056$ & 0.024 & 0.025 & 0.061 & 0.40 & 0.98 & --- \\
\bottomrule
\end{tabular}
\caption{Application estimates and bootstrap SEs at each primary-outcome cutpoint (200 refitted replicates). BiGGen uses 200 human ratings. SE ratios divide proposed SE by the indicated comparator's SE. BiGGen MSE ratios compare proposed to covariate-only adjustment over 500 repeated assignments against the fully rated prompt-pool reference; dashes indicate that no truth-based MSE is available for the other applications. IPW uses known probabilities in BiGGen and fitted treatment or labeling probabilities where needed in the other applications.}
\label{tab:empirical}
\end{table}
Across the 18 dataset--cutpoint combinations in Table~\ref{tab:empirical}, proposed bootstrap SEs are lower than IPW in all 18 (reductions of 5--64\%) and lower than covariate-only adjustment in 16 (reductions up to 28\%). SE gains depend on the cutpoint and comparator: proposed SE slightly exceeds covariate-only SE for BiGGen at $y=3$ and ACTG at $y=100$. In BiGGen's repeated-assignment benchmark (Appendix Table~\ref{tab:biggen-full}), the proposed estimator reduces MSE relative to covariate-only adjustment and performs comparably to thresholdwise PPI++; neither method dominates at every cutpoint. Our theoretical results additionally cover selective MAR labeling and uniform inference across outcome thresholds, settings beyond this MCAR, finite-grid benchmark.

\section{Discussion and Limitations}
A model update can affect different parts of the human-rating distribution in different ways. Distributional treatment effects make these changes visible without assigning numerical distances between rating categories. For ordinal outcomes, simultaneous inference across all nontrivial cutpoints describes uncertainty in the entire distribution contrast. This supports evaluation decisions that consider several quality thresholds together, rather than relying on an average score or selecting a favorable cutpoint.

The experiments show that surrogate information can improve precision beyond covariate adjustment, although the benefit varies across thresholds. In BiGGen's repeated-assignment benchmark, the proposed estimator reduces MSE relative to covariate-only adjustment and performs comparably to thresholdwise PPI++. These results support using automated ratings alongside a limited human-rating sample, while showing that the additional benefit depends on the information those ratings provide. The theoretical framework also accommodates selective labeling under MAR, extending beyond the randomized labeling used in BiGGen.

Finite-sample calibration remains an important follow-up. Ordinal simultaneous coverage is close to nominal in our simulation, whereas continuous-outcome bands slightly under-cover. Replacing fitted conditional CDFs with their exact values improves coverage, suggesting that outcome estimation deserves particular attention in further work. The asymptotic guarantees therefore do not settle how much labeled data is sufficient for reliable bands in a given application.

\subsection*{AI use statement}
Generative AI was used to review and revise the manuscript, discuss the statistical formulation and proofs, identify relevant literature, implement and check experimental procedures, generate figures and tables from computed results. Cached LLM-generated surrogate ratings are used in BiGGen, HelpSteer2, and wine; their construction is described in Appendix~\ref{app:experiments}. AI assistance does not replace author responsibility for the final mathematical claims, code, empirical results, or text.

\subsection*{Ethics statement}
This work uses simulations and secondary analyses of existing datasets. Human ratings serve as reference outcomes for specific evaluation rubrics and should not be interpreted as universal measures of quality. LLM-generated surrogate scores may reflect model and dataset biases; improved statistical efficiency does not itself establish fairness or causal validity. The empirical applications illustrate uncertainty estimation, and the ACTG~175 analysis is not intended to support clinical recommendations. Dataset sources and preprocessing are documented in Appendix~\ref{app:experiments} and the supplementary materials.

\subsection*{Reproducibility statement}
Algorithm~\ref{alg:dte} gives the estimation workflow; Appendices~\ref{app:algorithm}--\ref{app:experiments} provide implementation details, proofs, and experimental specification. The accompanying implementation saves replication-level estimates, standard errors, coverage indicators, bootstrap draws of empirical estimates, and the generated tables and figures. Judge scores for the experiments are also included in the reproduction package.

\bibliographystyle{iclr2027_conference}
\bibliography{references}
\appendix

Notation used in the main text and proofs is collected in Appendix~\ref{app:notation}.

\section{Implementation Details}
\label{app:algorithm}
Algorithm~\ref{alg:dte} uses inverse-label weights $1/\widehat\rho(X,w,S)$ when fitting $\widehat\mu_w$ on labeled training units under MAR. Bound fitted inverse weights as in Assumption~\ref{ass:rates}. If estimating $\mu_w$ by integrating $\widehat\nu_w$ over all arm-$w$ training units instead, use an additional training split or other conditions sufficient to establish the stated nuisance rates.

Train all propensity, labeling, outcome, and feature transformations outside the evaluation fold. Observed primary outcomes in that fold cannot be used for surrogate calibration or preprocessing in its training pipeline. In the experiments the surrogate scores are fixed existing measurements; no calibration to the primary-outcome scale is learned.

An all-zero or all-one threshold target in a training arm is represented by its constant fitted probability. For ordinal outcomes, the all-inclusive final category is omitted from studentization because its DTE is identically zero.

\clearpage
\section{Summary of Mathematical Notation}
\label{app:notation}
Arguments are suppressed when unambiguous. $\mathcal Y$ is the theoretical threshold index set; experiments evaluate fixed finite grids within it.

\begingroup
\small
\renewcommand{\arraystretch}{1.16}
\begin{tabular}{@{}p{.27\linewidth}p{.69\linewidth}@{}}
\toprule
Symbol & Definition or interpretation \\
\midrule
\multicolumn{2}{@{}l}{\textbf{Observations and targets}}\\
$O_i$, $n$ & Observed record $(X_i,W_i,S_i,R_i,R_iY_i)$; number of independent records. \\
$X$, $W$, $w$ & Pre-treatment covariates; binary arm assignment; an arm in $\{0,1\}$. \\
$Y$, $Y(w)$, $S$ & Primary outcome; potential primary outcome under arm $w$; surrogate observed for every unit. \\
$R$, $I_w$, $Z_y$ & Primary-outcome observation indicator; $\ind\{W=w\}$; $\ind\{Y\le y\}$. Empirical uses of the latent $Z_y$ are multiplied by $R$. \\
$\mathcal Y$, $\mathcal Y_0$, $J$ & Fixed threshold index set; subset used for studentized bands; number of thresholds in a finite-grid specialization. \\
$F_w(y)$, $\Delta(y)$ & $P\{Y(w)\le y\}$ and $F_1(y)-F_0(y)$. The standardized interpretation is $F_w(y)=\E_X\mu_w(y,X)$; the causal interpretation requires exchangeability. \\
\midrule
\multicolumn{2}{@{}l}{\textbf{Nuisance functions and estimation}}\\
$\pi_w(x)$, $\rho(x,w,s)$ & $P(W=w\mid X=x)$ and $P(R=1\mid X=x,W=w,S=s)$. \\
$\epsilon$, $\rho_{\min}$ & Positive population lower bounds for $\pi_w$ and $\rho$. \\
$\mu_w(y,x)$ & $P(Y\le y\mid X=x,W=w)$. \\
$\nu_w(y,x,s)$ & $P(Y\le y\mid X=x,W=w,S=s)$; under MAR, also the conditional CDF of the primary outcome among labeled records. \\
$\eta$, $\bar\eta$, $\widehat\eta^{-k}$ & True nuisance collection $(\mu,\nu,\pi,\rho)$; fixed candidate collection; collection fitted outside fold $k$. \\
$K$, $\mathcal I_k$, $n_k$, $k(i)$ & Number of folds; evaluation indices in fold $k$; $|\mathcal I_k|$; fold containing observation $i$. \\
$\Psi_w$, $\widehat\Psi_{wi}$ & Augmented arm score in \eqref{eq:score}; its evaluation at $O_i$ using $\widehat\eta^{-k(i)}$. \\
$\widehat F_w$, $\widehat\Delta$ & Average fitted arm score and difference between the two arm averages, as in \eqref{eq:estimate}. \\
\midrule
\multicolumn{2}{@{}l}{\textbf{Influence functions and inference}}\\
$\varphi_w(y;O)$, $\varphi(y;O)$ & $\Psi_w(y;O,\eta)-F_w(y)$ and $\varphi_1(y;O)-\varphi_0(y;O)$. \\
$\mathbb G$, $\Omega(y,y')$ & Tight Gaussian limit; covariance kernel $\E[\varphi(y;O)\varphi(y';O)]$. On a finite grid, $\boldsymbol\varphi=(\varphi(y_j;O))_j$ and $\Omega_{j\ell}=\Omega(y_j,y_\ell)$. \\
$\widehat\varphi_i(y)$, $\widehat\Omega(y,y')$ & Centered fitted contrast score; its empirical covariance kernel in \eqref{eq:se}. \\
$\sigma(y)$, $V(y)$ & $\sqrt{\Omega(y,y)}$; studentized limit $\mathbb G(y)/\sigma(y)$ on $\mathcal Y_0$. Hats denote plug-in versions. \\
$\zeta_i^{(b)}$, $T^{(b)}$, $\widehat c_{1-\alpha}$ & Rademacher multipliers; supremum statistic in \eqref{eq:band}; its conditional $(1-\alpha)$-quantile, approximated by the empirical quantile across multiplier draws. \\
\bottomrule
\end{tabular}
\endgroup

\paragraph{Proof conventions.}
$Pf=\int f\,dP$, $P_nf=n^{-1}\sum_i f(O_i)$, and $P_{n,k}f=n_k^{-1}\sum_{i\in\mathcal I_k}f(O_i)$. The training sigma-field is $\mathcal A_k$; $\|f\|_2=(P[f^2])^{1/2}$ holds its training sample fixed and is generally random for fitted $f$. In Appendix~\ref{app:joint}, $\delta_f=\widehat f-f$, $\xi_w=\nu_w-\mu_w$, and $\eta_w=Z_y-\nu_w$; this subscripted residual $\eta_w$ is distinct from the nuisance collection $\eta$. $\mathbb G_n=\sqrt n(P_n-P)$ and $\mathbb G_{n,k}=\sqrt{n_k}(P_{n,k}-P)$ are empirical processes. $N(u,\mathcal F,L^2(Q))$ is the covering number at radius $u$. $P_*,\E_*$ denote multiplier probability and expectation conditional on the data. In Appendix~\ref{app:efficiency}, $L^2_0(P)$ is the space of square-integrable centered functions, $\mathcal T$ is the observed-data tangent space, and $\mathcal T_X,\mathcal T_W,\mathcal T_S,\mathcal T_R,\mathcal T_Y$ are its orthogonal factor components.

\clearpage
\section{Proofs}
\label{app:proofs}
Throughout, $I_w=\ind\{W=w\}$, $Z_y=\ind\{Y\le y\}$, and all nuisance arguments are implicit when unambiguous. Statements involving latent $Z_y$ are expectations; its empirical contribution is always multiplied by~$R$.

\subsection{Identification and the score bias identity}
MAR gives $\E[Z_y\mid X,W=w,S,R=1]=\nu_w$. The tower property gives $\E[\nu_w\mid X,W=w]=\mu_w$, and consistency and exchangeability give $F_w=\E_X\mu_w$. Thus each arm CDF is identified from the observed-data law as
\[
 F_w(y)=\E_X\!\left[\E\!\left\{
 P(Y\le y\mid X,W=w,S,R=1)
 \mid X,W=w\right\}\right].
\]
The inner expectation averages over the surrogate distribution among all records in arm $w$ conditional on $X$, not only labeled records. Treatment and labeling positivity ensure that the required conditional distributions are identified on the target support.

For fixed candidate fits, conditional expectation of the final score correction is
\[
 \E\!\left[\frac{I_wR}{\bar\pi_w\bar\rho}(Z_y-\bar\nu_w)
 \middle| X,W,S\right]
 =\frac{I_w\rho}{\bar\pi_w\bar\rho}(\nu_w-\bar\nu_w).
\]
Combining with the second term yields
\[
 \E[\Psi_w(\bar\eta)]
 =\E\bar\mu_w+\E\!\left[\frac{I_w}{\bar\pi_w}(\nu_w-\bar\mu_w)\right]
 +\E\!\left[\frac{I_w}{\bar\pi_w}
 (\bar\nu_w-\nu_w)\frac{\bar\rho-\rho}{\bar\rho}\right].
\]
Condition the middle term on $X$ and subtract $F_w=\E\mu_w$. The first two terms become $\E[(\bar\mu_w-\mu_w)(\bar\pi_w-\pi_w)/\bar\pi_w]$, proving \eqref{eq:bias}. This also proves the asserted robustness combinations without presuming consistency of all four candidate functions.
At the true nuisance functions, the bias vanishes, giving the score representation of the identified targets:
\[
 F_w(y)=\E[\Psi_w(y;O,\eta)],\qquad
 \Delta(y)=\E[\Psi_1(y;O,\eta)-\Psi_0(y;O,\eta)].
\]

\subsection{Functional limit, covariance consistency, and uniform bands}
\label{app:joint}
We prove Theorem~\ref{thm:joint}. The partition $\{\mathcal I_k\}_{k=1}^K$ is fixed or drawn independently of the observations. Conditional on the partition, let $n_k=|\mathcal I_k|$, with $n_k/n\to\lambda_k\in(0,1)$, and define
\[
 \mathcal A_k=\sigma(O_i:i\notin\mathcal I_k),\qquad
 P_{n,k}f=\frac1{n_k}\sum_{i\in\mathcal I_k}f(O_i),\qquad
 P_nf=\frac1n\sum_{i=1}^n f(O_i).
\]
Any independent algorithmic randomization is included in $\mathcal A_k$. For an $\mathcal A_k$-measurable random function $f$, $Pf=\int f(o)\,dP(o)$ holds the training sample fixed; hence $\E[P_{n,k}f\mid\mathcal A_k]=Pf$. Each argument below conditions on one fold's training sample, not on all fitted nuisance functions simultaneously.

\paragraph{Step 0: entropy, measurability, and a local maximal inequality.}
All suprema below are measurable by Assumption~\ref{ass:rates}; equivalently, one may work with separable versions and outer probabilities. Choose conditional distributions as CDF kernels, so $y\mapsto\mu_w(y,x)$ and $y\mapsto\nu_w(y,x,s)$ are nondecreasing and right-continuous. Each class indexed by $y$ is bounded by one and has nested subgraphs. Such a class is VC-subgraph, as is $\{\ind\{Y\le y\}:y\in\mathcal Y\}$; restricting the index set preserves this property. The true influence function can be written
\[
 \varphi_w(y)=\left(1-\frac{I_w}{\pi_w}\right)\mu_w(y)
 +\frac{I_w}{\pi_w}\left(1-\frac R\rho\right)\nu_w(y)
 +\frac{I_wR}{\pi_w\rho}Z_y-F_w(y).
\]
The coefficients are fixed bounded functions of the observation, independent of $y$. For a bounded weight $a$, $\|a(f-g)\|_{Q,2}\le\|a\|_\infty\|f-g\|_{Q,2}$; finite sums admit covers obtained by multiplying the component covering numbers. Consequently the true arm-score and contrast-influence classes have constant envelopes and polynomial uniform covering numbers. The class of constants $\{F_w(y):y\in\mathcal Y\}$ has the same property. Their entropy integrals are finite, giving the $P$-Donsker property by the uniform-entropy CLT \citep{vandervaart1996weak}. This uses CDF ordering, not continuity of the outcome distribution or a density assumption. The resulting Gaussian process has sample paths uniformly continuous in the $L^2(P)$ semimetric; Euclidean continuity across outcome atoms is not asserted.

Condition on one training field $\mathcal A_k$. On the event in \eqref{eq:entropy}, the score-error class $\mathcal D_{kw}=\{d_{kw}(y):y\in\mathcal Y\}$, its squared class, and finite sums and products of bounded component classes inherit polynomial entropy with deterministic constants. Here one can cover the difference class by the Cartesian product of covers for the fitted and true score classes; no smoothness of the map $y\mapsto\widehat\eta(y)$ is assumed. Its bounded envelope depends only on the overlap and entropy constants. For any such class $\mathcal D$ on $m$ independent evaluation observations, write $s=\sup_{d\in\mathcal D}\|d\|_2$. The local maximal inequality of \citet[Thm.~2.1]{vandervaart2011local}, after rescaling the envelope, implies for $0<a<1$ and $s\le a$,
\begin{equation}
 \E\|\mathbb G_m\|_{\mathcal D}
 \le C\left\{a\sqrt{\log(C/a)}+\frac{\log(C/a)}{\sqrt m}\right\},
 \qquad \mathbb G_m=\sqrt m(P_m-P),
 \label{eq:local-maximal}
\end{equation}
with $C$ independent of the realized training data on that event. Indeed the uniform entropy integral is bounded by $C a\sqrt{\log(C/a)}$, and substitution into that theorem gives \eqref{eq:local-maximal}. Enlarging $C$ absorbs the constant envelope. If $s=o_P(1)$, conditional Markov's inequality first with fixed $a$, then $m\to\infty$, then $a\downarrow0$, proves $\|\mathbb G_m\|_{\mathcal D}=o_P(1)$. The exceptional entropy event has probability tending to zero. Thus no deterministic rate for $s$ beyond consistency is needed for this centered remainder.

For a fixed $J$-point grid, any fitted score class has at most $J$ elements, so \eqref{eq:entropy} holds without restricting its learner. For the implemented fixed-dimensional logistic regressions, conditional on training, standardization is an affine change of the fixed features, and each CDF fit belongs to the class of logistic transforms of linear predictors, including the two constant fits. This is a bounded VC-subgraph class. The fitted inverse weights are bounded and do not depend on $y$, so the preceding covering argument gives \eqref{eq:entropy} with constants depending on feature dimension and overlap, not on the training sample. This verifies complexity only: correct approximation and the uniform rates in \eqref{eq:rates} remain substantive requirements. For more flexible learners one must verify \eqref{eq:entropy} or supply an alternative vanishing local-entropy bound; cross-fitting alone is insufficient.

\paragraph{Step 1: exact score expansion and conditional bounds.}
Fix $k,w$ and suppress $(-k)$ on fitted nuisances. For each $f\in\{\mu_w,\nu_w,\pi_w,\rho\}$, write $\delta_f=\widehat f-f$. Also write $\xi_w=\nu_w-\mu_w$ and $\eta_w=Z_y-\nu_w$. The difference
$d_{kw}(y,O)=\Psi_w(y;O,\widehat\eta^{-k})-\Psi_w(y;O,\eta)$ decomposes exactly as $\sum_{a=1}^4 T_{a,kw}(y,O)$, where
\begin{align}
 T_{1,kw}&=\delta_{\mu_w}\left(1-\frac{I_w}{\widehat\pi_w}\right),&
 T_{2,kw}&=\frac{I_w\delta_{\nu_w}}{\widehat\pi_w}
                \left(1-\frac R{\widehat\rho}\right),\nonumber\\
 T_{3,kw}&=I_w\xi_w\left(\frac1{\widehat\pi_w}-\frac1{\pi_w}\right),&
 T_{4,kw}&=I_wR\eta_w\left(\frac1{\widehat\pi_w\widehat\rho}
                              -\frac1{\pi_w\rho}\right).
 \label{eq:four-errors}
\end{align}
The population residuals satisfy
\begin{equation}
 \E[\xi_w\mid X,W=w]=0,\qquad
 \E[\eta_w\mid X,W=w,S,R=1]=0.
 \label{eq:residual-centering}
\end{equation}
Consequently, conditioning successively on $X$, $(X,W,S)$, and $(X,W,S,R)$ gives
\begin{align}
 PT_{1,kw}&=P\left[\frac{\delta_{\mu_w}\delta_{\pi_w}}{\widehat\pi_w}\right],&
 PT_{2,kw}&=P\left[\frac{I_w\delta_{\nu_w}\delta_\rho}
                             {\widehat\pi_w\widehat\rho}\right],&
 PT_{3,kw}&=PT_{4,kw}=0.
 \label{eq:error-means}
\end{align}
These are conditional expectations with respect to a fresh evaluation record, with training fits held fixed. In particular, the exact bias identity and Cauchy--Schwarz imply
\begin{equation}
 |Pd_{kw}(y)|\le\frac2\epsilon\|\delta_{\mu_w}\|_2\|\delta_{\pi_w}\|_2
 +\frac4{\epsilon\rho_{\min}}\|\delta_{\nu_w}\|_2\|\delta_\rho\|_2.
 \label{eq:conditional-bias-bound}
\end{equation}
To control second moments, the overlap bounds and probability-valued estimates give
\begin{align*}
 \left|\frac1{\widehat\pi_w}-\frac1{\pi_w}\right|
 &\le\frac2{\epsilon^2}|\delta_{\pi_w}|,\\
 \left|\frac1{\widehat\pi_w\widehat\rho}-\frac1{\pi_w\rho}\right|
 &\le\frac4{\epsilon^2\rho_{\min}^2}
           (|\delta_{\pi_w}|+|\delta_\rho|).
\end{align*}
Using $|\xi_w|,|\eta_w|\le1$, we obtain the four explicit bounds
\begin{align*}
 \|T_{1,kw}\|_2&\le(1+2/\epsilon)\|\delta_{\mu_w}\|_2,\\
 \|T_{2,kw}\|_2&\le(2/\epsilon)(1+2/\rho_{\min})\|\delta_{\nu_w}\|_2,\\
 \|T_{3,kw}\|_2&\le(2/\epsilon^2)\|\delta_{\pi_w}\|_2,\\
 \|T_{4,kw}\|_2&\le\{4/(\epsilon^2\rho_{\min}^2)\}
                    (\|\delta_{\pi_w}\|_2+\|\delta_\rho\|_2).
\end{align*}
Thus, for a constant $C$ depending only on the overlap bounds,
\begin{equation}
 \sup_{y\in\mathcal Y}\|d_{kw}(y)\|_2
 \le C\sum_{f\in\{\mu_w,\nu_w,\pi_w,\rho\}}
        \sup_{y\in\mathcal Y}\|\widehat f-f\|_2=o_P(1).
 \label{eq:score-l2}
\end{equation}
Equation~\eqref{eq:conditional-bias-bound} and Assumption~\ref{ass:rates} separately yield $\sup_y|Pd_{kw}(y)|=o_P(n^{-1/2})$. Consistency controls the centered empirical error; the product rates control its conditional mean.

\paragraph{Step 2: uniform remainder and functional central limit theorem.}
Adding and subtracting the population score gives the exact expansion
\begin{align}
 \sqrt n\{\widehat F_w(y)-F_w(y)\}
 &=\mathbb G_n\varphi_w(y)+R_{1,nw}(y)+R_{2,nw}(y),\nonumber\\
 R_{1,nw}(y)&=\sum_{k=1}^K\sqrt{n_k/n}\,\mathbb G_{n,k}d_{kw}(y),\nonumber\\
 R_{2,nw}(y)&=\sqrt n\sum_{k=1}^K(n_k/n)Pd_{kw}(y),
 \qquad\mathbb G_{n,k}=\sqrt{n_k}(P_{n,k}-P).
 \label{eq:fold-expansion}
\end{align}
By \eqref{eq:score-l2}, $s_{kw}:=\sup_y\|d_{kw}(y)\|_2=o_P(1)$. To make the conditional-to-unconditional step explicit, let $E_{kw}$ be the training-measurable event on which the entropy bound holds. For fixed $t>0$ and $0<a<1$, \eqref{eq:local-maximal} gives
\begin{align*}
 P\{\|\mathbb G_{n,k}\|_{\mathcal D_{kw}}>t\}
 &\le P(E_{kw}^c)+P(s_{kw}>a)\\
 &\quad+\frac Ct\left\{a\sqrt{\log(C/a)}+n_k^{-1/2}\log(C/a)\right\}.
\end{align*}
First let $n\to\infty$ and then $a\downarrow0$. Since $K$ is fixed,
$\sup_y|R_{1,nw}(y)|=o_P(1)$. Equation~\eqref{eq:conditional-bias-bound} and the uniform product rates similarly give
\[
 \sup_y|R_{2,nw}(y)|\le\sqrt n\sum_k(n_k/n)\sup_y|Pd_{kw}(y)|=o_P(1).
\]
No independence between foldwise remainders is used. Subtracting arms yields
\begin{equation}
 \left\|\sqrt n(\widehat\Delta-\Delta)-\mathbb G_n\varphi\right\|_{\mathcal Y}
 =o_P(1),\qquad \|h\|_{\mathcal Y}:=\sup_{y\in\mathcal Y}|h(y)|.
 \label{eq:uniform-linearization}
\end{equation}
Step~0 shows that $\mathcal F_\varphi=\{\varphi(y;\cdot):y\in\mathcal Y\}$ is $P$-Donsker. Thus $\mathbb G_n\varphi\rightsquigarrow\mathbb G$ in $\ell^\infty(\mathcal Y)$, with covariance $P[\varphi(y)\varphi(y')]$ because $P\varphi(y)=0$. Functional Slutsky applied to \eqref{eq:uniform-linearization} proves the process limit, including asymptotic tightness. It also gives $\|\widehat\Delta-\Delta\|_{\mathcal Y}=O_P(n^{-1/2})$. All these statements hold at each fixed law $P$ satisfying the assumptions; uniformity over thresholds is distinct from uniformity over a class of data-generating laws.

\paragraph{Step 3: uniform covariance consistency and centering.}
Set $q_i(y)=\widehat\Psi_{1i}(y)-\widehat\Psi_{0i}(y)$,
$b_n(y)=\widehat\Delta(y)-\Delta(y)$, and
$e_i(y)=q_i(y)-\{\varphi(y;O_i)+\Delta(y)\}$. On fold $k$,
$e_i(y)=e_k(y,O_i):=d_{k1}(y,O_i)-d_{k0}(y,O_i)$.
The conditional squared-error class has a bounded envelope and polynomial entropy. Its uniform empirical-process expectation is bounded by a deterministic constant, using \eqref{eq:local-maximal} with a fixed radius after envelope rescaling. Consequently, conditional Markov and the vanishing probability of the exceptional entropy event give
\[
 \sup_y|(P_{n,k}-P)e_k(y)^2|=o_P(1).
\]
Since $\sup_yPe_k(y)^2=o_P(1)$ by Step~1, and there are finitely many folds,
\begin{equation}
 r_n^2:=\sup_{y\in\mathcal Y}\frac1n\sum_i e_i(y)^2=o_P(1).
 \label{eq:empirical-score-error}
\end{equation}
This uniform law replaces the finite union bound; pointwise conditional Markov bounds alone would not establish it.

Let $\widetilde\varphi_i(y)=q_i(y)-\Delta(y)=\varphi(y;O_i)+e_i(y)$ and let $C_\varphi$ bound the true influence functions. Expansion and Cauchy--Schwarz yield
\begin{equation}
 \sup_{y,y'\in\mathcal Y}\left|
 \frac1n\sum_i\widetilde\varphi_i(y)\widetilde\varphi_i(y')
 -\frac1n\sum_i\varphi(y;O_i)\varphi(y';O_i)\right|
 \le 2C_\varphi r_n+r_n^2=o_P(1).
 \label{eq:covariance-cross-terms}
\end{equation}
Empirical centering has the exact effect
\begin{equation}
 \widehat\Omega(y,y')=
 \frac1n\sum_i\widetilde\varphi_i(y)\widetilde\varphi_i(y')-b_n(y)b_n(y'),
 \qquad\sup_{y,y'}|b_n(y)b_n(y')|=O_P(n^{-1}).
 \label{eq:centering-correction}
\end{equation}
The bounded true-score product class is Glivenko--Cantelli: covers for each factor give polynomial entropy for their products. Its uniform LLN, \eqref{eq:covariance-cross-terms}, and \eqref{eq:centering-correction} prove
\begin{equation}
 \sup_{y,y'\in\mathcal Y}|\widehat\Omega(y,y')-\Omega(y,y')|=o_P(1).
 \label{eq:covariance-consistency}
\end{equation}
In particular, with $\sigma(y)=\sqrt{\Omega(y,y)}$ and $\widehat\sigma(y)=\sqrt{\widehat\Omega(y,y)}$, $\|\widehat\sigma-\sigma\|_{\mathcal Y}=o_P(1)$; on $\mathcal Y_0$ with $\inf_{\mathcal Y_0}\sigma>0$, also $\|1/\widehat\sigma-1/\sigma\|_{\mathcal Y_0}=o_P(1)$.

\paragraph{Step 4: conditional multiplier process convergence.}
Let $\zeta_i$ be independent Rademacher variables, independent of the observations, and define
\[
 G_n^*(y)=n^{-1/2}\sum_i\zeta_i\widehat\varphi_i(y),\qquad
 G_{n,0}^*(y)=n^{-1/2}\sum_i\zeta_i\varphi(y;O_i).
\]
For the true $P$-Donsker class, the conditional multiplier CLT \citep[Thm.~2.9.6]{vandervaart1996weak} gives $G_{n,0}^*\rightsquigarrow_*\mathbb G$ in $\ell^\infty(\mathcal Y)$, in probability. Applying the theorem to empirically centered true scores gives the same result: their difference is $(P_n\varphi)(n^{-1/2}\sum_i\zeta_i)$, whose supremum norm is $O_P(n^{-1/2})O_{P_*}(1)$.

We verify that replacing true scores by cross-fitted scores is conditionally negligible. On fold $k$, conditional on $\mathcal A_k$, the pairs $(O_i,\zeta_i)$ are i.i.d. under $P\otimes P_\zeta$. The class $\{(o,z)\mapsto z e_k(y,o):y\in\mathcal Y\}$ is centered, has the same $L^2$ radii as the error class, and inherits its entropy bound since $|z|=1$. Therefore \eqref{eq:local-maximal} and the fixed-radius argument in Step~2 give
\[
 H_{n,k}^*(y):=n_k^{-1/2}\sum_{i\in\mathcal I_k}\zeta_i e_k(y,O_i),
 \qquad\|H_{n,k}^*\|_{\mathcal Y}=o_P(1)
\]
under the joint law of data and multipliers. This also gives conditional negligibility in probability: for any $t,\delta>0$,
\[
 P\{P_*(\|H_{n,k}^*\|_{\mathcal Y}>t)>\delta\}
 \le\delta^{-1}P(\|H_{n,k}^*\|_{\mathcal Y}>t)\longrightarrow0,
\]
Finally,
\[
 G_n^*-G_{n,0}^*=
 \sum_k\sqrt{n_k/n}\,H_{n,k}^*-b_n\,n^{-1/2}\sum_i\zeta_i.
\]
The last term is conditionally negligible because $\|b_n\|_{\mathcal Y}=O_P(n^{-1/2})$ and $\E_*(n^{-1/2}\sum_i\zeta_i)^2=1$. Hence
\begin{equation}
 G_n^*\rightsquigarrow_*\mathbb G\quad\text{in }\ell^\infty(\mathcal Y),\ \text{in probability}.
 \label{eq:uniform-multiplier}
\end{equation}
Here conditional convergence means convergence, in probability, of conditional expectations against bounded Lipschitz functionals. We never condition simultaneously on all training fields to claim independence of fitted scores.

\paragraph{Step 5: studentization, quantiles, and coverage.}
On $\mathcal Y_0$, Steps~2--4 and continuity of multiplication by the bounded function $1/\sigma$ imply
\[
 \frac{\sqrt n(\widehat\Delta-\Delta)}{\widehat\sigma}
 \rightsquigarrow V:=\frac{\mathbb G}{\sigma},\qquad
 \frac{G_n^*}{\widehat\sigma}\rightsquigarrow_* V
 \quad\text{in }\ell^\infty(\mathcal Y_0).
\]
For the multiplier statement, the error in replacing $\widehat\sigma$ by $\sigma$ is bounded by $\|G_n^*\|_{\mathcal Y_0}\|1/\widehat\sigma-1/\sigma\|_{\mathcal Y_0}=o_{P_*}(1)$ in probability, using conditional tightness and Step~3. The supremum norm is a continuous functional. Let $H(t)=P(\|V\|_{\mathcal Y_0}\le t)$ and $c=H^{-1}(1-\alpha)$. The theorem's quantile condition means $H$ is continuous at $c$ and $H(c-\delta)<1-\alpha<H(c+\delta)$ for every sufficiently small $\delta>0$. Conditional distribution convergence at continuity points on either side of $c$ implies $\widehat c_{1-\alpha}\to_P c$. A sandwich argument then gives
\begin{equation}
 P\{\Delta(y)\in C_n(y)\text{ for every }y\in\mathcal Y_0\}
 =P\left\{\left\|\frac{\sqrt n(\widehat\Delta-\Delta)}{\widehat\sigma}\right\|_{\mathcal Y_0}
       \le\widehat c_{1-\alpha}\right\}\longrightarrow1-\alpha.
 \label{eq:uniform-coverage}
\end{equation}
Only quantile regularity at the selected level is needed; it is stated explicitly for a continuum rather than inferred from a finite-dimensional argument.

For a fixed finite $\mathcal Y_0$, every coordinate of $V$ has variance one. Thus $P(\max_j|V_j|=t)\le\sum_jP(|V_j|=t)=0$. If its correlation matrix has rank $r\ge1$, write $V=AU$ for $U\sim N_r(0,I_r)$ with $A$ of full column rank. The map $u\mapsto\|Au\|_\infty$ is a norm, and every shell $\{u:a<\|Au\|_\infty<b\}$, $0<a<b$, is nonempty and open. Its Gaussian probability is positive, proving strict increase of $H$ and hence the finite-grid specialization even when the covariance is singular. At finite $J$, one can alternatively use conditional Chebyshev and a finite union bound in Step~2, so no learner-complexity restriction is necessary.

For unstudentized bands, let $\widehat q_{1-\alpha}$ be the conditional quantile of $\|G_n^*\|_{\mathcal Y}$. If the distribution of $\|\mathbb G\|_{\mathcal Y}$ satisfies the same quantile condition, the identical argument gives coverage $1-\alpha$ for $[\widehat\Delta(y)\pm\widehat q_{1-\alpha}/\sqrt n]$ simultaneously over $\mathcal Y$, including $\mathbb R$ when the assumptions hold there. No variance lower bound is needed for this unstudentized construction. In either construction, replacing the exact conditional quantile by $B_n$ independent draws preserves coverage when $B_n\to\infty$; the fixed 500 draws used in the experiments are a numerical approximation.

A numerical grid $\mathcal Y_{0,n}$ approximates continuum studentization only if its supremum error is negligible. A sufficient condition is that it is asymptotically dense in the Gaussian $L^2$ semimetric on $\mathcal Y_0$, together with stochastic equicontinuity of both studentized processes in that metric; then their suprema on $\mathcal Y_{0,n}$ differ from those on $\mathcal Y_0$ by $o_P(1)$ (conditionally for multipliers). The experiments do not invoke such a discretization limit: their reported event is coverage at every point of the stated fixed grid. \qed

\subsection{Observed-data efficient influence function}
\label{app:efficiency}
We prove the efficiency assertion of Theorem~\ref{thm:efficiency} in the observed-data model. The bound is stated for every fixed finite collection of thresholds; it does not assert a separate infinite-dimensional minimax bound. Fix $w$ and $y$. All conditional densities below may be probability mass functions or Radon--Nikodym derivatives relative to appropriate dominating measures, so the argument includes discrete and mixed primary outcomes.

\paragraph{Step 1: observed likelihood and tangent space.}
Write $p_1(v\mid x,w,s)=p_{Y\mid X,W,S,R=1}(v\mid x,w,s)$. With the outcome coordinate present only when $r=1$, the observed likelihood is
\begin{equation}
 p_O(o)=p_X(x)p_{W\mid X}(w\mid x)p_{S\mid X,W}(s\mid x,w)
         p_{R\mid X,W,S}(r\mid x,w,s)\,p_1(v\mid x,w,s)^r.
 \label{eq:observed-likelihood}
\end{equation}
In particular, the final factor is the \emph{observed} conditional outcome distribution. Under MAR, its CDF satisfies
\begin{equation}
 \nu_w^{R=1}(y,x,s):=\int\ind\{v\le y\}p_1(v\mid x,w,s)\,dv
 =\nu_w(y,x,s).
 \label{eq:observed-cdf}
\end{equation}
Thus the CDF defining the outcome tangent direction is the same $\nu_w$ that identifies the target. MAR and exchangeability impose no additional equality restrictions on the observed likelihood: any observed law with overlap admits a MAR full-data extension and an exchangeable potential-outcome extension. The positive lower bounds ensure finite inverse weights and allow sufficiently small bounded tilts to remain in a model with overlap. We impose no restriction such as a treatment-invariant surrogate distribution.

Let $L^2_0(P)$ be the Hilbert space of square-integrable, mean-zero functions with inner product $\langle f,g\rangle=P(fg)$. Consider a dominated one-dimensional submodel $\{P_t:|t|<t_0\}$ through $P_0=P$, differentiable in quadratic mean, with score $\dot\ell_O$. Differentiating the factors of \eqref{eq:observed-likelihood} gives
\begin{equation}
 \dot\ell_O=\dot\ell_X+\dot\ell_W+\dot\ell_S+\dot\ell_R+R\dot\ell_Y.
 \label{eq:observed-scores}
\end{equation}
Normalization of each factor implies conditional mean zero given its conditioning variables; for the last factor, $\E[\dot\ell_Y\mid X,W,S,R=1]=0$. Define the closed component spaces
\begin{align*}
 \mathcal T_X&=\{a(X):Pa=0,\ Pa^2<\infty\},\\
 \mathcal T_W&=\{a(X,W):\E[a\mid X]=0,\ Pa^2<\infty\},\\
 \mathcal T_S&=\{a(X,W,S):\E[a\mid X,W]=0,\ Pa^2<\infty\},\\
 \mathcal T_R&=\{a(X,W,S,R):\E[a\mid X,W,S]=0,\ Pa^2<\infty\},\\
 \mathcal T_Y&=\{R a(Y,X,W,S):\E[a\mid X,W,S,R=1]=0,\ P(Ra^2)<\infty\}.
\end{align*}
For example, $a_X\in\mathcal T_X$ and $a_W\in\mathcal T_W$ satisfy $P(a_Xa_W)=P\{a_X\E[a_W\mid X]\}=0$. The same conditioning argument handles all pairs; $\E[Ra\mid X,W,S,R]=0$ handles the outcome component against the first four. The component spaces are therefore mutually orthogonal.

Bounded, conditionally centered functions generate factor submodels through tilts $p_t=p(1+t a)$; centering preserves normalization and $|t|\|a\|_\infty<1$ ensures nonnegativity. Truncation followed by conditional centering approximates square-integrable component functions by bounded ones. Consequently the tangent space is
\begin{equation}
 \mathcal T=\mathcal T_X\oplus\mathcal T_W\oplus\mathcal T_S\oplus\mathcal T_R\oplus\mathcal T_Y.
 \label{eq:tangent-sum}
\end{equation}
It is the nonparametric observed-data tangent space: for any $h\in L^2_0(P)$, successive conditional-expectation differences along $X$, $(X,W)$, $(X,W,S)$, and $(X,W,S,R)$ give the first four components. The remaining innovation vanishes when $R=0$, because no outcome is then observed, and belongs to $\mathcal T_Y$. This establishes the completeness of \eqref{eq:tangent-sum}, not only pairwise orthogonality.

\paragraph{Step 2: pathwise derivative of the target.}
Along the submodel, write
\begin{equation}
 F_{w,t}(y)=\int\ind\{v\le y\}
       p_{1,t}(v\mid x,w,s)p_{S,t}(s\mid x,w)p_{X,t}(x)\,dv\,ds\,dx.
 \label{eq:path-functional}
\end{equation}
Differentiating the three factors at $t=0$ gives
\begin{align}
 \dot F_w(y)
 &=\E[\mu_w(y,X)\dot\ell_X(X)]
   +\E_X\E[\nu_w(y,X,S)\dot\ell_S\mid X,W=w]\nonumber\\
 &\quad+\E_X\E[Z_y\dot\ell_Y\mid X,W=w].
 \label{eq:path-derivative}
\end{align}
The last expectation uses the MAR extension of $p_1$; equivalently, its inner integral is with respect to $p_1(\cdot\mid X,w,S)$ and its outer surrogate integral uses $p_S(\cdot\mid X,w)$. Differentiation is immediate for bounded factor tilts. The resulting functional is continuous in the observed score's $L^2(P)$ norm, by the bounded inverse weights below, and hence extends to the closure of regular scores. There is no derivative in the treatment or labeling factor. In particular, throughout \eqref{eq:path-derivative} the outer $X$ law is marginal. The identity used for standardization is
\begin{equation}
 \E\!\left[\frac{I_w}{\pi_w(X)}h(Y,X,W,S)\right]
 =\E_X\E[h(Y,X,W,S)\mid X,W=w],
 \label{eq:standardization-identity}
\end{equation}
which follows by conditioning first on $X$. It cannot generally be replaced by $\E[h\mid W=w]$.

\paragraph{Step 3: membership of the proposed gradient.}
Decompose $\varphi_w=M_w+N_w+L_w$, where
\[
 M_w=\mu_w-F_w,\qquad N_w=\frac{I_w}{\pi_w}(\nu_w-\mu_w),\qquad
 L_w=\frac{I_wR}{\pi_w\rho}(Z_y-\nu_w).
\]
All three are bounded, with $|M_w|\le1$, $|N_w|\le1/\epsilon$, and $|L_w|\le1/(\epsilon\rho_{\min})$. Since $P\mu_w=F_w$, $M_w\in\mathcal T_X$. Also,
\[
 \E[N_w\mid X,W]=\frac{I_w}{\pi_w}
             \{\E[\nu_w\mid X,W=w]-\mu_w\}=0,
\]
where the displayed expression is zero off the arm by $I_w=0$, and the tower property applies on the arm. Hence $N_w\in\mathcal T_S$.

For the outcome component take
\begin{equation}
 h_Y=\frac{I_w}{\pi_w\rho}(Z_y-\nu_w),\qquad
 \E[h_Y\mid X,W,S]=0.
 \label{eq:outcome-direction}
\end{equation}
On $\{W=w\}$, this conditional expectation equals
$(\pi_w\rho)^{-1}\E[Z_y-\nu_w\mid X,W=w,S]=0$; on $\{W\ne w\}$ it vanishes because $I_w\equiv0$. By \eqref{eq:observed-cdf}, the same two-case argument gives $\E[h_Y\mid X,W,S,R=1]=0$ for the observed outcome factor. Thus its observed-data score is $Rh_Y=L_w$, and $L_w\in\mathcal T_Y$.

To exhibit the submodel explicitly, perturb only the observed outcome factor:
\[
 p_{1,t}(v\mid x,u,s)=p_1(v\mid x,u,s)
 \left[1+t\frac{\ind\{u=w\}}{\pi_w(x)\rho(x,u,s)}
                   \{\ind\{v\le y\}-\nu_w(y,x,s)\}\right].
\]
Its conditional integral is one, and it is nonnegative for $|t|<\epsilon\rho_{\min}$. Boundedness permits local density tilts with observed score $Rh_Y$. Analogous tilts of $p_X$ by $M_w$ and of $p_S$ by $N_w$ realize the other two components. Therefore $\varphi_w\in\mathcal T_X\oplus\mathcal T_S\oplus\mathcal T_Y\subset\mathcal T$.

\paragraph{Step 4: derivative representation and the bound.}
Orthogonality reduces the derivative representation to three inner products. For the covariate factor,
\[
 P(\varphi_w\dot\ell_X)=P(M_w\dot\ell_X)=P(\mu_w\dot\ell_X),
\]
since $P\dot\ell_X=0$. For the surrogate factor, \eqref{eq:standardization-identity} and conditional centering give
\begin{align*}
 P(\varphi_w\dot\ell_S)
 &=P\!\left[\frac{I_w}{\pi_w}(\nu_w-\mu_w)\dot\ell_S\right]\\
 &=\E_X\E[\nu_w\dot\ell_S\mid X,W=w].
\end{align*}
For the outcome factor, $R^2=R$, MAR, and $\E[\dot\ell_Y\mid X,W,S,R=1]=0$ imply
\begin{align*}
 P(\varphi_w R\dot\ell_Y)
 &=P\!\left[\frac{I_wR}{\pi_w\rho}(Z_y-\nu_w)\dot\ell_Y\right]\\
 &=P\!\left[\frac{I_w}{\pi_w}
       \E[(Z_y-\nu_w)\dot\ell_Y\mid X,W,S,R=1]\right]\\
 &=\E_X\E[Z_y\dot\ell_Y\mid X,W=w].
\end{align*}
Finally $P(\varphi_w\dot\ell_W)=P(\varphi_w\dot\ell_R)=0$ by \eqref{eq:tangent-sum}. Combining these identities with \eqref{eq:path-derivative} proves
\begin{equation}
 \dot F_w(y)=P(\varphi_w\dot\ell_O)\qquad\text{for every regular submodel score.}
 \label{eq:gradient-representation}
\end{equation}
The representation extends to $\mathcal T$ by Cauchy--Schwarz. Together with membership in $\mathcal T$, it identifies $\varphi_w$ as the canonical gradient \citep{bickel1993efficient,tsiatis2006semiparametric}. Linearity makes $\varphi_1-\varphi_0$ the canonical gradient of $\Delta(y)$, and $\Omega$ the efficient covariance matrix for the finite target vector. For any regular asymptotically linear estimator with influence vector $g$, $g-\boldsymbol\varphi$ is orthogonal to $\mathcal T$ coordinatewise; hence
\[
 P(gg^\top)=\Omega+P[(g-\boldsymbol\varphi)(g-\boldsymbol\varphi)^\top]\succeq\Omega.
\]
The convolution theorem gives the corresponding lower bound for regular estimators more generally \citep[Thm.~25.20]{vandervaart2000asymptotic}. Identification of the canonical gradient above follows from its tangent-space membership and derivative representation; the convolution theorem supplies the lower bound once that gradient is identified.

\paragraph{Step 5: regularity and attainment.}
To make attainment precise, fix a differentiable-in-quadratic-mean submodel with scalar score $s$, information $I=Ps^2$, and local sequence $t_n=h/\sqrt n$. Local asymptotic normality \citep[Thm.~7.2]{vandervaart2000asymptotic} gives
\[
 \log\frac{dP_{t_n}^{\otimes n}}{dP^{\otimes n}}
 =\frac h{\sqrt n}\sum_i s(O_i)-\frac{h^2I}{2}+o_P(1).
\]
The joint CLT for $(\boldsymbol\varphi,s)$ has cross-covariance $P(\boldsymbol\varphi s)=\dot{\boldsymbol\Delta}$. Contiguity transfers the asymptotic-linear remainder of Appendix~\ref{app:joint} to $P_{t_n}^{\otimes n}$. Le Cam's third lemma \citep[Ex.~6.7]{vandervaart2000asymptotic} shifts the Gaussian mean of $n^{-1/2}\sum_i\boldsymbol\varphi(O_i)$ by $h\dot{\boldsymbol\Delta}$. Since
$\sqrt n(\boldsymbol\Delta_{t_n}-\boldsymbol\Delta)=h\dot{\boldsymbol\Delta}+o(1)$,
\[
 \sqrt n(\widehat{\boldsymbol\Delta}-\boldsymbol\Delta_{t_n})
 \rightsquigarrow N_J(0,\Omega)\qquad\text{under }P_{t_n}^{\otimes n}.
\]
The limit does not depend on $h$, establishing local regularity and attainment of the efficient covariance bound. If treatment or labeling probabilities are known, the corresponding score spaces are removed from \eqref{eq:tangent-sum}; the gradient remains in the reduced tangent space and the same conclusion follows. No stronger restriction on the surrogate law is used. \qed

\subsection{Variance decompositions}
\label{app:variance}
We prove the variance comparison in Theorem~\ref{thm:efficiency} and the surrogate-gain identity \eqref{eq:surrogategain}. Fix $y$ and evaluate all scores at the true nuisances. First,
\[
 PH_w=P\!\left[\frac{I_wRZ_y}{\pi_w\rho}\right]
 =P\!\left[\frac{I_w\nu_w}{\pi_w}\right]=P\mu_w=F_w.
\]
Expanding \eqref{eq:score} gives the exact difference
\begin{equation}
 H_w-F_w=\varphi_w+A_w+B_w,\qquad
 A_w=\mu_w(I_w/\pi_w-1),\quad B_w=I_w\nu_w(R/\rho-1)/\pi_w.
 \label{eq:oracle-difference}
\end{equation}
Let $\mathcal C=\sigma(X,W,S)$ and $\mathcal D=\sigma(X,W,S,R)$. The following conditional identities collect the needed orthogonality statements:
\begin{align*}
 &\E[A_w\mid X]=0,\qquad \E[B_w\mid\mathcal C]=0,\\
 &M_v\in\sigma(X),\quad \E[N_v\mid X,W]=0,\quad
 N_v\in\mathcal C,\quad \E[L_v\mid\mathcal D]=0.
\end{align*}
Here $M_v\in\sigma(X)$ denotes measurability. For either $v,w\in\{0,1\}$, these give all six cross-inner-products:
\begin{align*}
 P(M_vA_w)&=P\{M_v\E[A_w\mid X]\}=0,&
 P(M_vB_w)&=P\{M_v\E[B_w\mid\mathcal C]\}=0,\\
 P(N_vA_w)&=P\{A_w\E[N_v\mid X,W]\}=0,&
 P(N_vB_w)&=P\{N_v\E[B_w\mid\mathcal C]\}=0,\\
 P(L_vA_w)&=P\{A_w\E[L_v\mid\mathcal D]\}=0,&
 P(L_vB_w)&=P\{B_w\E[L_v\mid\mathcal D]\}=0.
\end{align*}
Therefore $P\{(\varphi_1-\varphi_0)(A_1-A_0+B_1-B_0)\}=0$. Taking variances in the arm difference of \eqref{eq:oracle-difference} proves \eqref{eq:ipwgain}. Moreover $\Cov(A_v,B_w)=0$ by conditioning on $\mathcal C$, so neither the treatment-weight nor label-weight residual can cancel the other in that variance gap. This proves the known-weight oracle comparison directly, without treating oracle IPW as the influence function of an estimator that fits its weights.

For the covariate-only outcome comparator, direct subtraction gives
\begin{equation}
 C_w-\Psi_w=D_w:=\frac{I_w}{\pi_w}\left(\frac R\rho-1\right)(\nu_w-\mu_w).
 \label{eq:ablation-difference}
\end{equation}
Here $\E[D_w\mid\mathcal C]=0$, so $PD_w=0$. The previous calculations with $D_w$ in place of $B_w$ yield $P(M_vD_w)=P(N_vD_w)=P(L_vD_w)=0$ for both arms. In addition $D_1D_0=0$ by disjoint arm support and thus $\Cov(D_1,D_0)=0$. As $R\mid\mathcal C$ is Bernoulli with parameter $\rho$,
\[
 \E[(R/\rho-1)^2\mid\mathcal C]
 =\frac{\rho}{\rho^2}-2\frac\rho\rho+1=\frac{1-\rho}{\rho}.
\]
Consequently
\begin{align*}
 \Var(C_1-C_0)-\Var(\varphi_1-\varphi_0)
 &=\Var(D_1-D_0)=\sum_{w=0}^1 PD_w^2\\
 &=\sum_{w=0}^1P\!\left[\frac{I_w}{\pi_w^2}\frac{1-\rho}{\rho}
                         (\nu_w-\mu_w)^2\right],
\end{align*}
which is \eqref{eq:surrogategain}. Since each integrand is nonnegative and $\pi_w,\rho$ are bounded away from zero, the gap is zero if and only if
$I_w(1-\rho)(\nu_w-\mu_w)^2=0$ almost surely for both arms. It is strictly positive if this quantity is positive with positive probability for at least one arm. The identities apply at each specified threshold; usefulness for a conditional mean alone does not imply a gain at every cutpoint. \qed

\section{Experimental Specification and Additional Results}
\label{app:experiments}
\paragraph{Implementation and recorded outputs.}
All manuscript figures and tables are embedded directly from saved replication-level results by the figure-generation script. Every simulation uses seeds $1000+b$ for repetition $b$; bootstrap multipliers use independent seeds $900000+b$. HelpSteer2 and wine masks use seed~0; application record bootstraps use seed~707. BiGGen assignment and annotation seeds are specified below. Folds are generated by shuffled five-fold splitting with seed~123. All empirical methods in a comparison share bootstrap resamples.

\paragraph{Conditional CDF learning.}
We use logistic regression with an intercept, $L^2$ regularization parameter \texttt{C=1}, a maximum of 1000 iterations, and standardization fitted within the relevant training data. $\mu_w$ uses inverse-label weights normalized to have training-sample mean one; $\nu_w$ uses $(X,S)$ without label weights. The ablation uses the same fitted $\mu_w$ for both prediction slots. Known probabilities are used for the simulation comparisons in Figure~\ref{fig:simulation}, empirical MCAR masks, and the ACTG working treatment probability. The 5000-repetition band experiment estimates the simulation probabilities: $\pi$ from $X$ and $\rho$ from $(X,W,S)$, using the same cross-fitting folds. Estimated $\rho$ also supplies the weights for fitting $\mu_w$. The logistic labeling model contains the constant ordinal MCAR probability but only approximates the bounded continuous MAR mechanism below. Unknown scores use logistic regression, with fitted probabilities clipped to $[0.02,0.98]$. Working outcome regressions need not equal the true conditional CDFs, even with the simulation's quadratic feature.

\subsection{Simulation design and diagnostics}
\label{app:simulation-details}

\paragraph{Oracle IPW normalization.}
The simulation's oracle IPW baseline uses known weights and the Horvitz--Thompson arm estimator $n^{-1}\sum_i I_{wi}R_iZ_{yi}/(\pi_w\rho_i)$. It does not divide by the realized sum of weights. Under MCAR with label probability $r$ and $\pi_0=\pi_1=1/2$, its exact DTE mean squared error is
\[
 \operatorname{MSE}\{\widehat\Delta_{\mathrm{HT}}(y)\}
 =\frac1n\left[\frac{2}{r}\{F_1(y)+F_0(y)\}-\Delta(y)^2\right].
\]
This follows from unbiasedness, disjoint arm indicators, and $Z_y^2=Z_y$. As both CDFs approach one, the MSE approaches $4/(nr)$ rather than zero: the estimated arm totals still fluctuate. Thus increasing right-tail MSE is expected for this unnormalized estimator. H\'ajek IPW instead divides each arm's weighted outcome total by its weight total; under randomized MCAR sampling it equals the labeled-only empirical CDF comparator. The variance guarantee in Theorem~\ref{thm:efficiency} concerns oracle IPW, and gains over it should not be interpreted as gains over normalized IPW.

\paragraph{Continuous MAR labeling.}
Let $f_W(X)$ be the arm-specific mean in Section~\ref{sec:simulation}, and $b(X)=0.3(X_1-0.5X_2)$. We set
\[
 \rho(X,W,S)=0.05+0.30\operatorname{expit}\{0.4X_1+0.4[S-f_W(X)-b(X)]\}.
\]
This is a fixed function of observed variables, bounded between $0.05$ and $0.35$. Since the logit argument is centered and symmetric under the DGP, $\E\rho=0.20$. Label selection depends on the surrogate and is MAR rather than MCAR. All simulated records remain i.i.d.; the probability is not normalized by a sample-dependent empirical mean. Continuous thresholds are $\{-0.75,-0.5,0,0.5,1,1.5\}$. At the left boundary, $F_0(-0.75)=0.1119$ and $F_1(-0.75)=0.1016$; both arms are evaluated at the same outcome threshold. Ordinal cutpoints are $\{0,1,2,3\}$ after assigning categories $0$--$4$ using the four stated boundaries.

\paragraph{Known truth and simulation summaries.}
The control CDF is $\Phi\{y/\sqrt{0.3^2+0.2^2+0.5^2}\}$. The treated CDF is
\[
 F_1(y)=\int_{-\infty}^{\infty}\Phi\!\left(\frac{y-0.4x^2}{\sqrt{0.6^2+0.5^2}}\right)\phi(x)\,dx.
\]
Numerical quadrature over $[-10,10]$ with absolute tolerance $10^{-11}$ supplies the reference. The ordinal CDFs equal these continuous CDFs at the category boundaries. Bias is the replication mean error, MSE is the replication mean squared error, and coverage is the fraction of intervals containing this reference. Reported SE calibration compares the mean estimated SE to the empirical standard deviation across repetitions. Neither bias nor coverage is inferred from a single real-data point estimate. Figure~\ref{fig:simulation} and the ordinal baseline comparisons use 500 repetitions at $n=2000$ with known probabilities. The band experiment uses 5000 repetitions for each outcome at $n=2000$, estimating both $\pi$ and $\rho$.

\paragraph{Diagnosing continuous-band undercoverage.}
\label{app:coverage-diagnostic}
We use 5000 paired repetitions at $n=2000$, replacing both fitted outcome CDFs with their exact DGP values. The $n=2000$ run uses the same data seeds as the main band experiment but retains known treatment and labeling probabilities to isolate outcome fitting. Writing $f_w(X)$ for the arm mean and $b(X)$ for the surrogate bias, Gaussian conditioning gives
\[
 \mu_w(y,X)=\Phi\!\left\{\frac{y-f_w(X)}{0.5}\right\},\qquad
 \nu_w(y,X,s)=\Phi\!\left\{\frac{y-f_w(X)-\frac{0.25}{0.61}[s-b(X)-f_w(X)]}
                                    {\sqrt{0.25\cdot0.36/0.61}}\right\}.
\]
Treatment and labeling probabilities, evaluation grids, and multiplier seeds are unchanged. With $B=500$ Rademacher draws, replacing the fitted conditional CDFs by their true functions raises coverage from $0.929$ to $0.944$, a paired gain of $0.015$ (Monte Carlo SE $0.003$). Increasing to $B=5000$, retaining the first 500 draws, yields coverage $0.931$ with fitted CDFs and $0.944$ with exact CDFs. Thus outcome fitting contributes to undercoverage, but replacing it does not fully restore nominal coverage.

Misses are dispersed across cutpoints, with the largest coordinate miss rate ($0.015$) shared by $y=-0.75$ and $y=-0.5$; mean estimated SE divided by empirical SD ranges from $0.981$ to $1.007$. Thus average SE calibration can coexist with a poorly calibrated maximum studentized error. Since design weights are known, the conditional bias identity is zero for arbitrary training-fold outcome regressions whenever the fits exist. This diagnostic implicates the fitted-score approximation, but does not distinguish limited training data from the working logistic link. Supplementary outputs also report conditional Gaussian calibration and a separate population-covariance check.

\paragraph{Thresholdwise PPI baseline.}
For arm $w$, let $Q_i(y)=\ind\{S_i\le y\}$. Under MCAR, the baseline estimates its primary-outcome CDF as the average of $Q_i(y)$ over all arm-$w$ records plus the average of $Z_{iy}-Q_i(y)$ over labeled arm-$w$ records. These are nested samples. Writing their respective empirical means as $\bar Q_w$ and $\bar B_w$, its estimated influence function is
\[
 \frac{I_{wi}}{\widehat p_w}(Q_i-\bar Q_w)
 +\frac{I_{wi}R_i}{\widehat p_{wR}}(Z_{iy}-Q_i-\bar B_w),
\]
where $\widehat p_w=n_w/n$ and $\widehat p_{wR}=n_{wR}/n$. We compute the SE of the difference of these arm functions on the same records. This accounts for overlap between labeled and full samples; an independent-sample PPI variance formula would be inappropriate. No power-tuning parameter is fitted.

\begin{table}[t]
 \centering\small
\begin{tabular}{llrrrrr}
\toprule
Outcome & $y$ & Bias & SD & Mean SE & Mean SE / SD & Coverage \\
\midrule
Continuous & -0.75 & -0.0015 & 0.0283 & 0.0285 & 1.00 & 0.950 \\
 & -0.5 & -0.0028 & 0.0335 & 0.0332 & 0.99 & 0.944 \\
 & 0 & +0.0015 & 0.0379 & 0.0375 & 0.99 & 0.952 \\
 & 0.5 & +0.0014 & 0.0329 & 0.0326 & 0.99 & 0.944 \\
 & 1 & -0.0014 & 0.0232 & 0.0237 & 1.02 & 0.954 \\
 & 1.5 & -0.0022 & 0.0157 & 0.0159 & 1.01 & 0.942 \\
\midrule
Ordinal & 0 & -0.0013 & 0.0257 & 0.0265 & 1.03 & 0.958 \\
 & 1 & -0.0006 & 0.0362 & 0.0352 & 0.97 & 0.962 \\
 & 2 & +0.0021 & 0.0368 & 0.0367 & 1.00 & 0.962 \\
 & 3 & +0.0003 & 0.0308 & 0.0306 & 0.99 & 0.958 \\
\bottomrule
\end{tabular}
 \caption{Proposed-estimator pointwise diagnostics at $n=2000$ (500 repetitions). SD is the empirical standard deviation across repetitions; mean SE averages the score-based DTE SE retaining arm covariance.}
 \label{tab:calibration}
\end{table}

\subsection{Ordinal baseline implementations and paired comparisons}
\label{app:baseline-methods}
All added comparisons use the same 500 ordinal datasets at $n=2000$, generated with seeds $1000+b$. Their outcome-observation masks, targets, and cutpoints coincide with the original simulation. The algorithms below use only observed primary outcomes. They require randomized treatment and MCAR labeling and are not presented as solutions to selective labeling or observational treatment assignment.

\paragraph{Labeled-only regression-adjusted DTE.}
We implement the regression-adjusted CDF in \citet[Eq.~(1)]{byambadalai2024estimating} on the labeled subset $\mathcal L=\{i:R_i=1\}$. With $m=|\mathcal L|$, $m_w=\sum_{i\in\mathcal L}I_{wi}$, and five-fold cross-fitted logistic CDF predictions $\widehat g_{wi}(y)$ based only on $X$, the estimator is
\[
 \widehat F_w^{\mathrm{labeled\text{-}RA}}(y)
 =\frac1m\sum_{i\in\mathcal L}\widehat g_{wi}(y)
  +\frac1{m_w}\sum_{i\in\mathcal L}I_{wi}\{Z_{yi}-\widehat g_{wi}(y)\}.
\]
The nuisance learner and covariate features match the proposed method. Unlabeled covariates and all surrogate scores are excluded. The centered arm influence estimates are
\[
 \widehat g_{wi}-\overline g_w
 +\frac{I_{wi}}{m_w/m}\{Z_{yi}-\widehat g_{wi}-\overline e_w\},
\]
where $\overline g_w$ averages predictions over $\mathcal L$ and $\overline e_w$ averages residuals over its arm-$w$ records. We take the arm difference before estimating its variance and scale its SE by $\sqrt m$.

\paragraph{Thresholdwise PPI++.}
For each arm and threshold, write $q_y$ for a prediction of $Z_y$. Raw-score PPI++ uses $q_y=\ind\{S\le y\}$; learned-CDF PPI++ uses logistic regression on $(X,S)$, with the same learner as $\widehat\nu_w$. For outer fold $k$, estimate
\[
 \widehat\lambda_{kwy}
 =\operatorname{clip}_{[0,1]}\!\left[
 \widehat u_{kw}\,
 \frac{\widehat{\Cov}_{kw}(Z_y,q_y)}{\widehat{\Var}_{kw}(q_y)}\right],
\]
where $\widehat u_{kw}$ is the unlabeled fraction in the training arm. Covariance and variance use its labeled records; learned predictions for tuning come from an inner three-fold split. A zero proxy variance sets $\widehat\lambda=0$. The learned predictor is then refitted on the outer training arm's labels and evaluated on its held-out records. This is the disjoint-sample mean power-tuning formula of \citet[Sec.~6]{angelopoulos2023ppipp}, applied to threshold indicators; the factor $\widehat u_{kw}$ accounts for the finite unlabeled sample.

Let $v_{wi}=\widehat\lambda_{k(i)wy}\widehat q_{wi}(y)$, and let $u_w$ be the number of unlabeled arm-$w$ records. We compute
\[
 \widehat F_w^{++}(y)
 =\frac1{m_w}\sum_{i:W_i=w,R_i=1}(Z_{yi}-v_{wi})
  +\frac1{u_w}\sum_{i:W_i=w,R_i=0}v_{wi}.
\]
For SEs and bands, its centered influence estimates on all $n$ records are
\[
 \frac{I_{wi}R_i}{m_w/n}(Z_{yi}-v_{wi}-\overline a_w)
 +\frac{I_{wi}(1-R_i)}{u_w/n}(v_{wi}-\overline b_w),
\]
where $\overline a_w$ and $\overline b_w$ are the two corresponding sample averages. The arm difference retains their joint covariance. Tuning and preprocessing exclude the outer evaluation fold. Ordinary PPI retains the original nested-sample construction with unit weight; PPI++ uses disjoint samples and its matching tuning and SE formulas.

\paragraph{Evaluation.}
All new methods receive 500 shared Rademacher draws per repetition. Labeled-only RA uses the same draws restricted to labeled records. We include every prespecified cutpoint in the band event, including zero-SE cases. Table~\ref{tab:ordinal-baselines} reports empirical MSE, SE calibration, pointwise coverage, and simultaneous coverage and width.

\begin{table}[t]
\centering\small
\begin{tabular}{lrrrrrr}
\toprule
Method & MSE $\times10^3$ & Reduction (\%) & SE/SD & Pointwise coverage & Band & Width \\
\midrule
Proposed & 0.663--1.355 & 0 & 1.00 & 0.958--0.962 & 0.946 & 0.159 \\
Covariate-only & 0.755--1.656 & 8--21 & 0.99 & 0.944--0.960 & --- & --- \\
PPI++ (raw) & 0.732--1.767 & 9--23 & 1.00 & 0.938--0.958 & 0.958 & 0.173 \\
PPI++ (CDF) & 0.688--1.573 & 4--15 & 1.00 & 0.946--0.960 & 0.936 & 0.167 \\
Labeled-only RA & 0.814--1.728 & 18--24 & 0.99 & 0.936--0.960 & 0.942 & 0.177 \\
Threshold PPI & 1.276--2.174 & 34--48 & 0.97 & 0.934--0.944 & --- & --- \\
Labeled only & 0.946--2.448 & 30--45 & 0.99 & 0.940--0.956 & --- & --- \\
Oracle IPW & 1.055--7.880 & 37--88 & 0.99 & 0.938--0.950 & --- & --- \\
Surrogate only & 0.470--5.492 & $-177$--$82$ & 1.02 & 0.076--0.956 & --- & --- \\
\bottomrule
\end{tabular}
\caption{Paired ordinal MCAR comparisons: $n=2000$, 500 repetitions. MSE, reduction, and pointwise coverage report ranges across the four cutpoints. Reduction is $100(1-\mathrm{MSE}_{\mathrm{proposed}}/\mathrm{MSE}_{\mathrm{row}})$ at each cutpoint; negative values indicate higher proposed MSE. SE/SD averages the ratio of mean reported SE to empirical SD across cutpoints. Band is simultaneous 95\% coverage and Width is the mean full band width. Dashes indicate bands not computed for the simpler baselines. Monte Carlo SE at coverage 0.95 is approximately 0.010.}
\label{tab:ordinal-baselines}
\end{table}

\subsection{Additional application details}
\label{app:application-details}
\paragraph{Shared estimation and uncertainty.}
For each application, Table~\ref{tab:empirical} compares the proposed method, covariate-only outcome adjustment, and IPW on the same records and cutpoints. All use the five-fold nuisance-learning procedure specified above. Each reported SE is the standard deviation of 200 nonparametric record-bootstrap estimates, using seed~707 and common resamples across methods. Every replicate refits estimated nuisances and repeats cross-fitting; known design probabilities remain fixed. The observed assignment and labeling indicators are resampled with their records, rather than redrawn. This includes the fitting variability of estimated IPW weights. Results are reported separately at each cutpoint; score-based proposed SEs are saved separately from bootstrap SEs. These calculations do not adjust for additional prompt, reviewer, finite-population, or trial-design dependence. They describe empirical precision; population coverage is assessed in the simulations.

\subsubsection{BiGGen Bench}
\label{app:biggen}
\paragraph{Data, contrast, and surrogate.}
We use the public \texttt{human\_eval} split of BiGGen Bench \citep{kim2025biggen}; the pinned source revision is recorded in the reproduction package. It contains 2,780 responses from four models on 695 prompts. The primary contrast assigns $W=0$ to Mistral-7B-Instruct-v0.2 and $W=1$ to Mixtral-8x7B-Instruct-v0.1. The primary outcome is the released integer human rating on the 1--5 task-specific rubric; the surrogate is the cached GPT-4-Turbo-2024-04-09 score. We evaluate all nontrivial cutpoints $y\in\{1,2,3,4\}$ without rounding or calibrating human ratings. Four missing human ratings, encoded as $-1$, belong to Llama-2-13b-hf, which is excluded from the reported comparisons. The models used below have complete human, GPT-4 Turbo, and Claude ratings.

\paragraph{Labeling and covariates.}
Each prompt contributes one response, selected by independent Bernoulli$(1/2)$ model assignment; exactly 200 selected responses are then sampled uniformly for annotation. The fixed application uses repetition $b=0$ from \texttt{SeedSequence([20260920,b])}, without searching across assignments. Known probabilities are $\pi_w=1/2$ and marginal $\rho=200/695$. Covariates are log character and word counts for both user and system prompts, user-prompt character diversity, and seven capability indicators. Response features do not enter the treatment model. Unobserved primary outcomes are represented by zero before fitting; counterfactual ratings are withheld from every estimator. Using one response per prompt also prevents its two model outputs from entering different folds.

\paragraph{Estimation and uncertainty.}
The fixed application resamples its 695 observed prompt records under the shared 200-replicate bootstrap above. Both probabilities remain known; scaling and outcome fitting occur within training folds. In addition to the three main comparators, Table~\ref{tab:biggen-full} reports labeled-only, labeled-only RA, raw-score and learned-CDF PPI++, thresholdwise PPI, and surrogate-only estimates. Their implementations use the same learners and sampling conventions described in Appendix~\ref{app:baseline-methods}. Bootstrap SEs for the fixed application are distinct from the score-based SEs and empirical SDs in the repeated-assignment experiment below.

\paragraph{Results and interpretation.}
The target compares generation policies on released responses, not a deployed user A/B test. Rubrics vary by task, so ratings represent task-specific levels rather than a universal quality scale. BiGGen's annotator qualification used agreement with GPT-4-1106 scores; human--surrogate agreement is therefore not wholly independent validation. Cached scores are reused without judge training or new API calls. Table~\ref{tab:biggen-full} separates fixed-application precision from error against the fully rated benchmark. Differences depend on the cutpoint and comparator and do not establish uniform finite-sample superiority over PPI++.

\paragraph{Additional benchmark comparisons.}
For repetitions $b=0,\ldots,499$, \texttt{SeedSequence([20260920,b])} generates the assignment and 200-label sample. Assignments and labels are common across baselines and surrogate sensitivities. For the fixed prompt pool $\mathcal P$ and the two released outputs, the reference DTE is
\[
 \Delta_{\mathcal P}(y)=\frac{1}{|\mathcal P|}\sum_{i\in\mathcal P}
 \bigl[\ind\{Y_i(1)\le y\}-\ind\{Y_i(0)\le y\}\bigr].
\]
At each cutpoint separately, we average squared error against this reference over 500 repetitions. No MSE or SE is pooled across cutpoints. The uncertainty calculation uses the i.i.d. arm-score difference and 500 common Rademacher draws from \texttt{SeedSequence([20260920,b,1])}. Labeled-only RA restricts these draws to labeled units and uses its labeled sample size. Labeling is without replacement and the prompt pool is fixed; the reported SEs do not include finite-population corrections.

\begin{table}[t]
\centering\small
\begin{tabular}{llrrrrrr}
\toprule
& & \multicolumn{2}{c}{Fixed assignment} & \multicolumn{4}{c}{Repeated assignment and labeling} \\
\cmidrule(lr){3-4}\cmidrule(lr){5-8}
Method & $y$ & $\widehat\Delta$ & Boot. SE & MSE & SD & Mean SE & Coverage \\
\midrule
Proposed & 1 & $+0.025$ & 0.042 & 1.430 & 0.038 & 0.042 & 0.980 \\
 & 2 & $+0.012$ & 0.055 & 2.746 & 0.052 & 0.054 & 0.952 \\
 & 3 & $-0.058$ & 0.066 & 3.332 & 0.058 & 0.062 & 0.964 \\
 & 4 & $-0.052$ & 0.068 & 4.205 & 0.065 & 0.067 & 0.960 \\
\midrule
Covariate-only & 1 & $+0.023$ & 0.045 & 1.793 & 0.042 & 0.045 & 0.960 \\
 & 2 & $-0.002$ & 0.056 & 3.421 & 0.058 & 0.060 & 0.952 \\
 & 3 & $-0.081$ & 0.064 & 4.272 & 0.065 & 0.069 & 0.950 \\
 & 4 & $-0.071$ & 0.069 & 4.696 & 0.069 & 0.070 & 0.958 \\
\midrule
Labeled only & 1 & $+0.009$ & 0.043 & 1.654 & 0.041 & 0.043 & 0.952 \\
 & 2 & $-0.023$ & 0.057 & 3.288 & 0.057 & 0.058 & 0.950 \\
 & 3 & $-0.105$ & 0.065 & 4.269 & 0.065 & 0.067 & 0.952 \\
 & 4 & $-0.042$ & 0.070 & 4.414 & 0.066 & 0.068 & 0.958 \\
\midrule
Labeled-only RA & 1 & $+0.024$ & 0.045 & 1.751 & 0.042 & 0.044 & 0.952 \\
 & 2 & $+0.001$ & 0.057 & 3.332 & 0.057 & 0.059 & 0.956 \\
 & 3 & $-0.090$ & 0.066 & 4.100 & 0.064 & 0.067 & 0.950 \\
 & 4 & $-0.048$ & 0.068 & 4.479 & 0.067 & 0.069 & 0.960 \\
\midrule
PPI++ (raw) & 1 & $+0.029$ & 0.042 & 1.345 & 0.037 & 0.040 & 0.960 \\
 & 2 & $-0.045$ & 0.055 & 2.570 & 0.051 & 0.053 & 0.956 \\
 & 3 & $-0.090$ & 0.063 & 3.352 & 0.058 & 0.061 & 0.958 \\
 & 4 & $-0.020$ & 0.071 & 4.311 & 0.066 & 0.067 & 0.952 \\
\midrule
PPI++ (CDF) & 1 & $+0.018$ & 0.044 & 1.435 & 0.038 & 0.042 & 0.958 \\
 & 2 & $-0.006$ & 0.054 & 2.718 & 0.052 & 0.055 & 0.942 \\
 & 3 & $-0.074$ & 0.063 & 3.386 & 0.058 & 0.062 & 0.964 \\
 & 4 & $-0.051$ & 0.068 & 4.381 & 0.066 & 0.067 & 0.954 \\
\midrule
Threshold PPI & 1 & $+0.028$ & 0.043 & 1.453 & 0.038 & 0.041 & 0.968 \\
 & 2 & $-0.052$ & 0.057 & 2.855 & 0.053 & 0.055 & 0.958 \\
 & 3 & $-0.079$ & 0.069 & 3.844 & 0.062 & 0.065 & 0.962 \\
 & 4 & $+0.005$ & 0.080 & 5.034 & 0.071 & 0.073 & 0.962 \\
\midrule
Oracle IPW & 1 & $-0.020$ & 0.045 & 1.815 & 0.043 & 0.045 & 0.950 \\
 & 2 & $-0.080$ & 0.063 & 3.917 & 0.063 & 0.065 & 0.962 \\
 & 3 & $-0.190$ & 0.080 & 6.186 & 0.079 & 0.082 & 0.958 \\
 & 4 & $-0.200$ & 0.112 & 11.109 & 0.105 & 0.112 & 0.974 \\
\midrule
Surrogate only & 1 & $-0.026$ & 0.021 & 0.378 & 0.016 & 0.019 & 0.952 \\
 & 2 & $-0.055$ & 0.028 & 0.560 & 0.023 & 0.027 & 0.984 \\
 & 3 & $-0.025$ & 0.035 & 1.441 & 0.028 & 0.035 & 0.942 \\
 & 4 & $-0.006$ & 0.030 & 3.503 & 0.025 & 0.029 & 0.566 \\
\bottomrule
\end{tabular}
\caption{Pointwise BiGGen results with 200 human ratings. The first two numerical columns describe the fixed application in Table~\ref{tab:empirical}, including its refitted-bootstrap SE. Remaining columns use 500 repeated assignments and annotation samples: MSE ($\times10^{-3}$), empirical SD, mean score-based SE, and pointwise benchmark coverage, each computed separately at the indicated cutpoint. Surrogate-only targets automated ratings, not the human-rating distribution.}
\label{tab:biggen-full}
\end{table}

The model sensitivity replaces Mixtral with gpt-3.5-turbo-0125 on the same 695 prompts; the surrogate sensitivity replaces GPT-4 Turbo with Claude-3-Opus in the primary comparison. These choices were fixed before the 500-repetition runs. No repetition or cutpoint is removed, and all three 200-label conditions finish without an empty-arm fitting failure or zero-SE cutpoint. Table~\ref{tab:biggen-sensitivity} reports the additional comparisons. The submission contains the three 200-label conditions; exploratory 100-label runs remain outside the submission package.

\begin{table}[t]
\centering\small
\begin{tabular}{llrrrrrrr}
\toprule
Setting & $y$ & MSE & Mean SE & / Cov. & / RA & / PPI++ & / CDF & Coverage \\
\midrule
GPT-3.5 comparison & 1 & 1.443 & 0.040 & 0.842 & 0.883 & 1.005 & 0.989 & 0.960 \\
 & 2 & 2.789 & 0.054 & 0.799 & 0.802 & 0.967 & 0.967 & 0.952 \\
 & 3 & 3.342 & 0.062 & 0.763 & 0.782 & 0.993 & 0.978 & 0.950 \\
 & 4 & 3.922 & 0.067 & 0.860 & 0.885 & 0.960 & 0.928 & 0.962 \\
\midrule
Claude surrogate & 1 & 1.509 & 0.042 & 0.842 & 0.862 & 1.070 & 1.024 & 0.958 \\
 & 2 & 2.809 & 0.056 & 0.821 & 0.843 & 1.016 & 0.994 & 0.958 \\
 & 3 & 3.343 & 0.063 & 0.783 & 0.815 & 0.959 & 0.964 & 0.964 \\
 & 4 & 4.439 & 0.068 & 0.945 & 0.991 & 1.009 & 0.992 & 0.962 \\
\bottomrule
\end{tabular}
\caption{Pointwise BiGGen sensitivity results with 200 human ratings. Each row reports proposed MSE ($\times10^{-3}$), mean score-based SE, four MSE ratios against the indicated baselines, and pointwise benchmark coverage over 500 repetitions. GPT-3.5 replaces Mixtral as the treatment model; Claude replaces GPT-4 Turbo as the surrogate. No metric is averaged across cutpoints.}
\label{tab:biggen-sensitivity}
\end{table}

\subsubsection{HelpSteer2}
\label{app:helpsteer}
\paragraph{Data, contrast, and surrogate.}
We use the 20,324 prompt--response pairs in the HelpSteer2 training split \citep{wang2024helpsteer2}; the pinned source revision is recorded in the reproduction package. The primary outcome is released human helpfulness on the 0--4 scale. The contrast indicator is $W=\ind\{\text{complexity}\ge2\}$. The surrogate is the stored \texttt{gpt-5.6-terra} batch rating under the HelpSteer2 helpfulness rubric, without retraining on released human ratings. We evaluate $y\in\{0,1,2,3\}$. Missing judge scores trigger an error rather than imputation based on human ratings; the exact rubric and prompt template are included in the reproduction package.

\paragraph{Labeling and covariates.}
A Bernoulli MCAR mask with probability $0.02$ and NumPy seed~0 retains 399 human ratings. Hidden labels are replaced by zero before fitting. Covariates are $\log(1+\text{prompt characters})$, $\log(1+\text{prompt words})$, and distinct-character count divided by prompt length. The prepared record order and mask are fixed across methods.

\paragraph{Estimation and uncertainty.}
We estimate $\pi(X)$ by cross-fitted logistic regression and use the known labeling probability $\rho=0.02$. Proposed, covariate-only, and IPW estimates use the shared 200-replicate record bootstrap, refitting propensity and outcome regressions as applicable. Table~\ref{tab:empirical} reports each cutpoint's estimate and bootstrap SE; Figure~\ref{fig:empirical} visualizes the same contrasts with pointwise SE bars.

\paragraph{Results and interpretation.}
Complexity is an observed response attribute, so the target is a descriptive contrast standardized over prompt covariates rather than the causal effect of a model or prompt intervention. Responses may share prompts; record-bootstrap SEs are not cluster-adjusted deployment uncertainty. We do not claim absence of pretraining contamination in the judge. Smaller SEs alone establish neither accuracy nor coverage.

\subsubsection{Wine reviews}
\label{app:wine}
\paragraph{Data, contrast, and surrogate.}
The prepared sample contains 10,000 US wine reviews from \texttt{spawn99/wine-reviews}, filtered to nonmissing points, price, description, variety, and state, and restricted to unambiguously red or white varieties \citep{ji2025predictions}. We set $W=1$ for red and $W=0$ for white wine. The primary outcome is the professional points rating; the surrogate is the cached \texttt{gpt-5.6-terra} rating of the tasting description on a 1--5 rubric. We use raw judge ratings without calibration on professional ratings, evaluating primary-outcome thresholds $y\in\{85,87,89,91,93\}$. The prompt template and score order are recorded in the reproduction package.

\paragraph{Labeling and covariates.}
A Bernoulli MCAR mask with probability $0.20$ and NumPy seed~0 retains 2,049 professional ratings. Unobserved primary outcomes are represented by zero before fitting. Covariates are log price, its square, and indicators for California, Washington, Oregon, New York, and Virginia, with other states as the reference.

\paragraph{Estimation and uncertainty.}
We estimate $\pi(X)$ by cross-fitted logistic regression and use known $\rho=0.20$. Conditional regressions predict primary-outcome threshold probabilities from the auxiliary score directly, allowing the primary-outcome and surrogate scales to differ. Proposed, covariate-only, and IPW SEs use the shared 200-replicate record bootstrap, refitting all estimated nuisances. Table~\ref{tab:empirical} and Figure~\ref{fig:empirical} report the threshold-specific results.

\paragraph{Results and interpretation.}
Price and state standardization defines a descriptive target, not a manipulable wine-color intervention. Description-based surrogate ratings are auxiliary measurements of existing reviews rather than blinded assessments of new wines. The record bootstrap does not account for reviewer or winery dependence. As in HelpSteer2, these results compare precision without known population truth for validating coverage.

\subsubsection{ACTG 175}
\label{app:actg}
\paragraph{Data, contrast, and surrogate.}
We retain arms~0 and~1 of the released ACTG~175 data \citep{hammer1996trial}: 1,054 patients comparing zidovudine--didanosine ($W=1$) with zidovudine monotherapy ($W=0$). The primary outcome is week-96 CD4 (\texttt{cd496}), the surrogate is week-20 CD4 (\texttt{cd420}), and observation of week-96 CD4 is recorded by the supplied \texttt{r} indicator. We evaluate thresholds $y\in\{100,200,300,400,500\}$.

\paragraph{Labeling and covariates.}
There are 654 observed primary outcomes; no synthetic annotation mask is imposed. The baseline covariates are \texttt{cd40}, \texttt{age}, \texttt{wtkg}, \texttt{karnof}, \texttt{hemo}, \texttt{homo}, \texttt{drugs}, \texttt{race}, \texttt{gender}, \texttt{symptom}, \texttt{oprior}, \texttt{z30}, \texttt{preanti}, and \texttt{str2}. Missing primary-outcome values are masked before estimation, while week-20 CD4 enters the labeling and surrogate-augmented outcome regressions.

\begin{figure}[t]
 \centering
\definecolor{cproposed}{HTML}{1670b5}
\definecolor{cnosurrogate}{HTML}{c88400}
\definecolor{cppipp}{HTML}{cc6677}
\definecolor{cppipplearned}{HTML}{332288}
\definecolor{cgoldra}{HTML}{117733}
\definecolor{cppi}{HTML}{555555}
\definecolor{cgold}{HTML}{259642}
\definecolor{cipw}{HTML}{8b4fa6}
\definecolor{csurrogate}{HTML}{d24b40}
\begin{tikzpicture}
\begin{groupplot}[group style={group size=3 by 1,horizontal sep=.95cm},
scale only axis,width=.25\linewidth,height=3.35cm,
tick label style={font=\fontsize{8}{9}\selectfont},
label style={font=\fontsize{9}{10}\selectfont},title style={font=\small},
xlabel={Outcome cutpoint},grid=major,grid style={gray!15},
tick align=outside,major tick length=2pt,every axis plot/.append style={mark size=1.5pt},
legend style={font=\fontsize{8}{10}\selectfont,draw=none,column sep=12pt},legend columns=3]
\nextgroupplot[title={HelpSteer2},ylabel={CDF contrast},xtick={0,1,2,3}]
\addplot+[color=cproposed,mark=*,solid,line width=1pt,error bars/.cd,y dir=both,y explicit,error bar style={line width=.5pt}] coordinates {(-0.1,-0.05363401) +- (0,0.05370209) (0.9,-0.09645907) +- (0,0.07122981) (1.9,-0.1932248) +- (0,0.09177218) (2.9,-0.03723564) +- (0,0.09808214)};
\addplot+[color=cnosurrogate,mark=square*,dashed,line width=.7pt,error bars/.cd,y dir=both,y explicit,error bar style={line width=.5pt}] coordinates {(0,-0.05754857) +- (0,0.05907536) (1,-0.1087617) +- (0,0.07981642) (2,-0.2022917) +- (0,0.09600351) (3,-0.04629171) +- (0,0.1013081)};
\addplot+[color=cipw,mark=triangle,solid,line width=.7pt,error bars/.cd,y dir=both,y explicit,error bar style={line width=.5pt}] coordinates {(0.1,-0.05916902) +- (0,0.06345178) (1.1,-0.1162436) +- (0,0.08925467) (2.1,-0.2078907) +- (0,0.1077065) (3.1,-0.06681472) +- (0,0.1424985)};
\addplot[black,densely dotted,forget plot] coordinates {(0,0) (3,0)};
\nextgroupplot[title={Wine},xtick={85,89,93}]
\addplot+[color=cproposed,mark=*,solid,line width=1pt,error bars/.cd,y dir=both,y explicit,error bar style={line width=.5pt}] coordinates {(84.8,0.07267449) +- (0,0.02407251) (86.8,0.1057843) +- (0,0.03479824) (88.8,0.09814981) +- (0,0.03760439) (90.8,0.06767276) +- (0,0.03934799) (92.8,0.0622222) +- (0,0.0347629)};
\addplot+[color=cnosurrogate,mark=square*,dashed,line width=.7pt,error bars/.cd,y dir=both,y explicit,error bar style={line width=.5pt}] coordinates {(85,0.06501701) +- (0,0.0333302) (87,0.09448999) +- (0,0.04283571) (89,0.08857428) +- (0,0.04327376) (91,0.07048458) +- (0,0.04109123) (93,0.07822883) +- (0,0.03489816)};
\addplot+[color=cipw,mark=triangle,solid,line width=.7pt,error bars/.cd,y dir=both,y explicit,error bar style={line width=.5pt}] coordinates {(85.2,0.05367216) +- (0,0.04134502) (87.2,0.05227148) +- (0,0.05945751) (89.2,0.0241494) +- (0,0.07272455) (91.2,-0.01912456) +- (0,0.08588387) (93.2,-0.0372188) +- (0,0.09731982)};
\addplot[black,densely dotted,forget plot] coordinates {(85,0) (93,0)};
\nextgroupplot[title={ACTG 175},legend to name=empiricallegend,xtick={100,300,500}]
\addplot+[color=cproposed,mark=*,solid,line width=1pt,error bars/.cd,y dir=both,y explicit,error bar style={line width=.5pt}] coordinates {(90,-0.1067203) +- (0,0.05320404) (190,-0.1732889) +- (0,0.0621371) (290,-0.1528075) +- (0,0.06527325) (390,-0.118597) +- (0,0.0626545) (490,-0.05613871) +- (0,0.0477251)};
\addlegendentry{Proposed}
\addplot+[color=cnosurrogate,mark=square*,dashed,line width=.7pt,error bars/.cd,y dir=both,y explicit,error bar style={line width=.5pt}] coordinates {(100,-0.1035492) +- (0,0.05306333) (200,-0.1657682) +- (0,0.06425515) (300,-0.1428953) +- (0,0.06847971) (400,-0.1114982) +- (0,0.0630576) (500,-0.05763368) +- (0,0.04890174)};
\addlegendentry{Covariate-only}
\addplot+[color=cipw,mark=triangle,solid,line width=.7pt,error bars/.cd,y dir=both,y explicit,error bar style={line width=.5pt}] coordinates {(110,-0.09494187) +- (0,0.06179053) (210,-0.1426229) +- (0,0.08247597) (310,-0.115457) +- (0,0.1017759) (410,-0.09387817) +- (0,0.1137574) (510,-0.05690204) +- (0,0.1196649)};
\addlegendentry{IPW}
\addplot[black,densely dotted,forget plot] coordinates {(100,0) (500,0)};
\end{groupplot}
\node[anchor=north] at ([yshift=-1cm]group c2r1.south) {\pgfplotslegendfromname{empiricallegend}};
\end{tikzpicture}
 \caption{Empirical standardized distributional contrasts with $\pm1.96$ bootstrap-SE bars. Methods are horizontally offset around each primary-outcome cutpoint for visibility; all estimates and SEs are evaluated at the original cutpoints. The same nuisance-refitting bootstrap is used for proposed, covariate-only outcome adjustment, and IPW. Bars visualize estimated precision; they are not validated coverage statements or simultaneous bands.}
 \label{fig:empirical}
\end{figure}
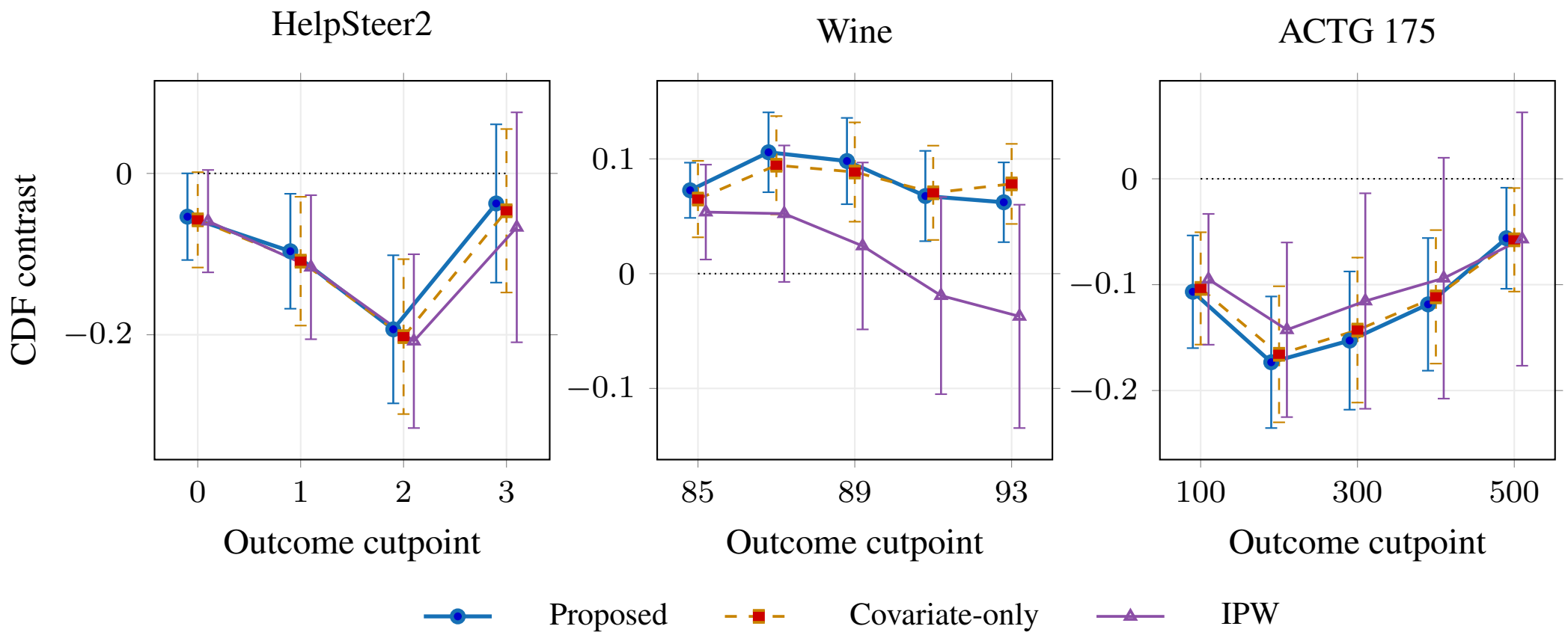

\paragraph{Estimation and uncertainty.}
We use the working treatment probability $\pi=1/2$ and estimate $\rho(X,W,S)$ by cross-fitted logistic regression. Under the stated MAR assumption, this permits label observation to depend on baseline covariates, treatment, and week-20 CD4. The shared 200-replicate record bootstrap refits labeling and outcome models. Table~\ref{tab:empirical} and Figure~\ref{fig:empirical} report the results at each CD4 threshold.

\paragraph{Results and interpretation.}
A fitted labeling model is not evidence of MAR or noninformative censoring. The i.i.d. working analysis does not reproduce the trial's original stratified assignment, so these bootstrap SEs are not a coverage claim under that design. No clinical decision or general stochastic-dominance conclusion is inferred from the illustrative estimates.

\end{document}